\documentclass[preprint,aps,nofootinbib,a4paper,superscriptaddress,longbibliography,amsfonts,amssymb,amsmath,titlepage]{revtex4-1}

\usepackage{bm}
\usepackage{bbm}
\usepackage{color}
\usepackage{microtype}
\usepackage{IEEEtrantools}
\usepackage{mathrsfs}
\usepackage{amsthm}
\usepackage{enumitem}
\usepackage[normalem]{ulem}

\usepackage{graphicx,color}
\usepackage{array}
\usepackage{tikz}
\usetikzlibrary{shapes.geometric}
\usetikzlibrary{decorations.pathmorphing}
\usetikzlibrary{arrows,shapes,positioning}
\usetikzlibrary{decorations.markings}
\tikzstyle arrowstyle=[scale=1]
\tikzstyle directed=[postaction={decorate,decoration={markings,mark=at position .65 with {\arrow[arrowstyle]{stealth}}}}]
\tikzstyle reverse directed=[postaction={decorate,decoration={markings,mark=at position .65 with {\arrowreversed[arrowstyle]{stealth};}}}]
\usepackage[pdftex]{hyperref}

\usepackage[T1]{fontenc}

\def\GHPwt{\ensuremath{\circeq}}
\newcommand{\GHPLie}{\text{\L}}

\def\sib{\bar\sigma}

\def\kab{\bar\kappa}

\def\rhb{\bar\rho}

\def\tab{\bar\tau}

\def\kap{\kappa'}
\def\sip{\sigma'}
\def\rhp{\rho'}
\def\tap{\tau'}

\font\ec=ecrm0800 at 12pt
\def\th{\hbox{\ec\char'336}}
\def\edth{\hbox{\ec\char'360}}
\def\thp{\hbox{\ec\char'336}'}
\def\edthp{\hbox{\ec\char'360}'}
\def\Dbar{\tilde \edth}
\def\Nbar{\tilde \th'}
\def\mb{\bar m}

\def\pheq{\phantom{=}}

\newcommand{\RR}{{\mathbb R}}
\newcommand{\CC}{{\mathbb C}}

\newtheorem{theorem}{Theorem}

\newtheorem{definition}{Definition}

\renewcommand{\S}{{\mathcal S}}
\newcommand{\E}{{\mathcal E}}
\newcommand{\T}{{\mathcal T}}
\newcommand{\Lie}{{\mathcal L}}
\renewcommand{\O}{{\mathcal O}}
\renewcommand{\P}{{\mathcal P}}

\newcommand{\sM}{{\mathscr M}}
\newcommand{\sS}{{\mathscr S}}
\newcommand{\sI}{{\mathscr I}}
\newcommand{\sH}{{\mathscr H}}
\newcommand{\sP}{{\mathscr P}}
\newcommand{\sL}{{\mathscr L}}
\newcommand{\sV}{{\mathscr V}}
\newcommand{\sW}{{\mathscr W}}
\newcommand{\sY}{{\mathscr Y}}
\newcommand{\sE}{{\mathscr E}}
\newcommand{\sX}{{\mathscr X}}
\newcommand{\sU}{{\mathscr U}}

\def\ben{\begin{equation}}
\def\een{\end{equation}}
\def\bena{\begin{eqnarray}}
\def\eena{\end{eqnarray}}

\def\dd{{\rm d}}
\def\half{\tfrac{1}{2}}
\renewcommand{\Re}{\operatorname{Re}}
\def\pd{\partial}

\begin{document}

\title{Teukolsky formalism for nonlinear Kerr perturbations}

\author{Stephen R. Green}
\email{stephen.green@aei.mpg.de}
\affiliation{Max Planck Institute for Gravitational Physics (Albert Einstein Institute)\\
  Am M\"uhlenberg 1, 14476 Potsdam, Germany}

\author{Stefan Hollands}
\email{stefan.hollands@uni-leipzig.de}
\affiliation{Institut f\"ur Theoretische Physik, Universit\"at Leipzig\\
  Br\"uderstrasse 16, D-04103 Leipzig, Germany}

\author{Peter Zimmerman}
\email{peter.zimmerman@aei.mpg.de}
\affiliation{Max Planck Institute for Gravitational Physics (Albert Einstein Institute)\\
  Am M\"uhlenberg 1, 14476 Potsdam, Germany}

\begin{abstract}
  We develop a formalism to treat higher order (nonlinear) metric
  perturbations of the Kerr spacetime in a Teukolsky framework. We
  first show that solutions to the linearized Einstein equation with
  nonvanishing stress tensor can be decomposed into a pure gauge part
  plus a zero mode (infinitesimal perturbation of the mass and spin)
  plus a perturbation arising from a certain scalar (``Debye-Hertz'')
  potential, plus a so-called ``corrector tensor.'' The scalar
  potential is a solution to the spin $-2$ Teukolsky equation with a
  source. This source, as well as the tetrad components of the
  corrector tensor, are obtained by solving certain decoupled ordinary
  differential equations involving the stress tensor. As we show,
  solving these ordinary differential equations reduces simply to
  integrations in the coordinate $r$ in outgoing Kerr-Newman
  coordinates, so in this sense, the problem is reduced to the
  Teukolsky equation with source, which can be treated by a separation
  of variables ansatz.  Since higher order perturbations are subject
  to a linearized Einstein equation with a stress tensor obtained from
  the lower order perturbations, our method also applies iteratively
  to the higher order metric perturbations, and could thus be used to
  analyze the nonlinear coupling of perturbations in the near-extremal
  Kerr spacetime, where weakly turbulent behavior has been conjectured
  to occur. Our method could also be applied to the study of
  perturbations generated by a pointlike body traveling on a timelike
  geodesic in Kerr, which is relevant to the extreme mass ratio
  inspiral problem.
\end{abstract}

\maketitle
\tableofcontents

\section{Introduction}
\label{sec:1}

Metrics describing a perturbed Kerr black hole play an important role
in gravitational physics, for instance in order to model the ringdown
phase after a black hole merger, or an extreme mass ratio
inspiral. Their analysis also presents nontrivial mathematical
challenges: one may wish to prove the expected decay properties of the
perturbations (see e.g.,
\cite{Dafermos:2016uzj,Dafermos:2017yrz,Andersson:2019dwi} and
references for recent progress directly relevant to this work), or to
understand the complicated but highly special geometrical structure of
the perturbation equations on the Kerr background (see e.g., the
monograph \cite{Chandrasekhar:1983}).

In the case of linear perturbations (i.e., solutions to the linearized Einstein equation around Kerr), one has the Teukolsky formalism \cite{Teukolsky:1973ha, TeukolskyLett1972}, 
which in effect reduces the problem to the study of 
a linear scalar wave equation. Furthermore, this equation can be solved by a separation of variables ansatz. Combined, these two features dramatically simplify the 
analysis of linear perturbations. As originally derived, Teukolsky's equation applies to the extreme 
components $\psi_0$ or $\psi_4$ of the linearized Weyl tensor, cf.~section \ref{sec:2}. 
However, it was discovered soon afterwards \cite{Chrzanowski:1975wv,Kegeles:1979an} that the equation also applies directly to the 
metric perturbation itself in a sense. In fact, if one makes an ansatz of the form $h_{ab} = \Re\S_{ab}^\dagger \Phi$, where $\S_{ab}^\dagger$ 
is a certain second order differential operator [cf.~\eqref{eq:Sdag}] constructed using geometrical objects of the background Kerr geometry, then if the complex ``Hertz potential'' $\Phi$ is a solution to the source-free (adjoint) Teukolsky equation, $h_{ab}$ is a solution to the source-free linearized Einstein equation. Furthermore, it is generally accepted---and argued more carefully in section \ref{sec:3}---that this ansatz in a sense covers all solutions with the exception only of pure gauge ones and ``zero modes,'' by which one means linear perturbations towards a Kerr black hole with different mass and/or spin. 

While the Weyl tensor components are sufficient to extract the information about gravitational radiation at null infinity, it is also highly desirable to know directly the metric perturbation itself. For instance, in the self-force approach \cite{Quinn:1996am,Mino:1996nk} to extreme mass ratio inspiral, the first step is to determine the linearized metric perturbation resulting from a stress tensor of a point particle on a geodesic in the Kerr background. If one is interested in higher order perturbations, one has to solve, for the $n$-th order perturbed metric, the linearized Einstein equation with a source determined by the metric perturbations up to order $(n-1)$. Both of these problems involve solving the linearized Einstein equation with a nonvanishing source. But in the presence of a source, the Hertz-potential ansatz $h_{ab} = \Re\S_{ab}^\dagger \Phi$ is generically no longer consistent! Furthermore, while the metric perturbation can be extracted from the perturbed Weyl scalars $\psi_0$ or $\psi_4$ for perturbations satisfying the source free linearized Einstein equation by means of the Teukolsky-Starobinsky relations, this is no longer true for the Einstein equation with source. Thus, it would seem that the Teukolsky formalism is not useful to study higher order perturbations of Kerr (for example), and the mileage gained from the highly special properties of the background seems lost again. 

In this paper, we propose a way around this problem, which takes advantage of the very special geometric properties of Kerr (and in fact, algebraically special solutions). 
Our starting point is the linearized Einstein equation with source\footnote{To lighten the notation, we absorb the conventional prefactor of 
$8\pi G$ into $T_{ab}$.}, 
\ben
\label{eq:LEinstein}
(\E h)_{ab} = T_{ab}, 
\een
where $\E$ is the linearized Einstein operator on the Kerr background \eqref{eq:E}.
In applications, $T_{ab}$ would be e.g., the stress tensor of a point source on a geodesic in Kerr, or the nonlinear terms in the equation for the 
$n$-th order gravitational perturbation. Instead of the Hertz-potential ansatz, which is generically inconsistent with a nontrivial source $T_{ab}$, we will argue that the metric 
perturbation can be decomposed in the following manner\footnote{To be precise, this decomposition holds under certain technical assumptions detailed in the main text. 
The essentially only unproven hypothesis is the absence of ``purely outgoing algebraically special perturbations'' with real frequency. This restriction, however, plays no
role for the main application we have in mind which is a retarded solution generated by a compact source.} 
(see section \ref{sec:2} for the precise definition of the operators $\S, \O, \T$)
\begin{equation}
  \label{eq:decintro}
  h_{ab} = (\Lie_\xi g)_{ab} + \dot g_{ab} + x_{ab} + \Re (\S^\dagger \Phi)_{ab} , 
\end{equation}
where:
\begin{itemize}
\item $(\Lie_\xi g)_{ab} = \nabla_a \xi_b + \nabla_b \xi_a$ is a pure gauge perturbation (irrelevant). 
\item $\dot g_{ab} = \frac{\dd}{\dd s} g_{ab}^{M(s),a(s)} |_{s=0}$ is a zero mode, i.e., perturbation to another metric in the Kerr family $g_{ab}^{M,a}$, where $(M(0),a(0))=(M,a)$ are the mass/spin parameter of the given Kerr background $(\sM,g_{ab})$.
\item $x_{ab}$ is called a {\em corrector} field. As we will show, its nontrivial tetrad components [cf.~\eqref{eq:xdef}] are obtained from $T_{ab}$ by solving three decoupled {\em ordinary} differential equations [cf.~\eqref{eq:tr1}]
along a congruence of outgoing null geodesics aligned with a principal null direction.  
\item $\Phi$ is a complex potential which is a solution to the adjoint Teukolsky equation with a source, 
\ben 
\label{eq:OJ}
\O^\dagger \Phi = \eta. 
\een
The source $\eta$ is determined in terms of $T_{ab}$ by an 
equation of the form $(\T^\dagger \eta)_{ab} = T_{ab} - (\E x)_{ab}$. As we will show, 
this equation can  be solved 
by integrating one {\em ordinary} differential equation along a congruence of outgoing null geodesics aligned 
with a principal null direction [cf.~\eqref{eq:tr2}, \eqref{eq:S1s}].
\end{itemize}

The difficult step in our algorithm is thus to solve the adjoint Teukolsky equation with source $\O^\dagger \Phi = \eta$, a scalar wave equation, which can be dealt with by a separation of variables ansatz~\cite{TeukolskyLett1972}. By contrast, the determination of the source $\eta$ and the tetrad components of the 
corrector field $x_{ab}$ involves the comparatively easy task of solving scalar {\em ordinary} differential equations
depending on $T_{ab}$. Thus, our method represents a drastic simplification compared to the nonseparated tensor equation  \eqref{eq:LEinstein}, taking advantage to the maximum possible extent of the special geometric features of Kerr. Before presenting the details of our algorithm in the following sections, let us mention three potential applications of our method, which were the motivation of our analysis. 

\subsection{Point sources}

\begin{figure}
  \includegraphics[width=0.6\textwidth,]{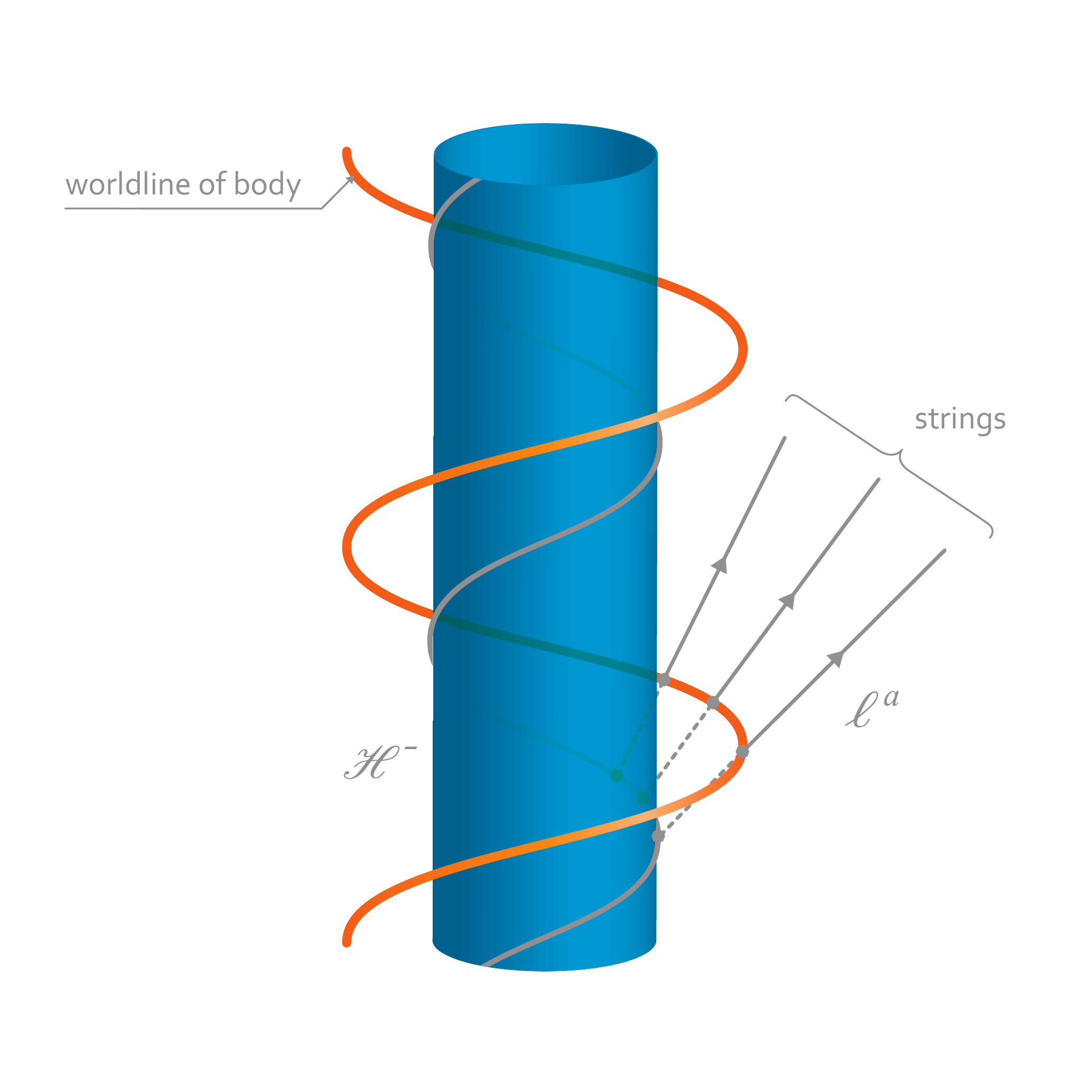}
  \caption{\label{fig:strings} Worldline of a body orbiting a Kerr
    black hole. The corrector tensor $x_{ab}$ is supported on the
    semi-infinite strings extending from the worldline to infinity.}
\end{figure}

Point sources for the linearized Einstein equation are described by a distributional stress tensor supported on a timelike geodesic $\gamma$, 
\ben
\label{eq:Tpoint}
T_{ab}(x) = m \int_{-\infty}^\infty  \dot \gamma_a(t) \dot \gamma_b(t) \delta(x-\gamma(t))\, \dd t \ .
\een
Since our corrector field $x_{ab}$ is obtained by integrating linear ordinary differential equations sourced by appropriate components of $T_{ab}$ along a congruence of outward geodesics from the past horizon to future null infinity, the corrector will only be nonzero for those geodesics intersecting $\gamma$. Furthermore, if we impose zero boundary conditions on $x_{ab}$ on $\mathscr{H}^-$, it follows that $x_{ab}$ is supported in each spatial slice along a ``string'' attached to the point source going out to infinity. The locus of this string is where the null geodesics intersecting $\gamma$ pierce the slice (see figure~\ref{fig:strings}). 
Thus, our analysis shows that the perturbation $h_{ab}$ is defined in terms of a Hertz potential (as well as possibly trivial zero mode and gauge perturbations), {\em plus} a corrector 
$x_{ab}$ supported on a semi-infinite string emanating at the point particle going to infinity. If instead vanishing boundary conditions on $x_{ab}$ are imposed at $\mathscr{I}^+$ rather than $\mathscr{H}^-$, then the string goes to the horizon rather than to infinity. 

The occurrence of such ``Dirac-type strings'' has been observed also by a number of previous authors \cite{Barack:2001ph} who have studied the equation  \eqref{eq:LEinstein} for point sources by a somewhat different approach. In their approach, the exterior of the black hole is divided into two domains $\sM_+$ and $\sM_-$ separated by an interface $\sS$ containing the particle orbit. In each domain, 
an ansatz $h^\pm_{ab} = \Re (\S^\dagger \Phi^\pm)_{ab}$ is made for the metric, and then the existence of a distributional source is encoded into suitable junction conditions 
relating $\Phi^+$ to $\Phi^-$ on the interface $\sS$ and in particular at the worldline $\gamma$ in the form of differentiability and continuity requirements. The perturbations obtained in this manner can be categorized \cite{Pound:2013faa} based on the form of the singularity into the ``half-string'' classes and the ``full-string'' class.  A ``no-string'' gauge may be formed by joining together the two regular sides of two opposite half-string perturbations along the interface $\sS$ through the particle. Such a perturbation, introduced by Friedman and collaborators in \cite{Keidl:2006wk,Keidl:2010pm}, avoids the stringlike singularities, but has instead a gauge discontinuity (and also delta-function distributions \cite{Pound:2013faa}) on the interface $\sS$.

Our construction scheme by contrast avoids the consideration of an interface, but the metric perturbation involves the corrector piece $x_{ab}$, which has a delta-function-like 
distribution along a string emanating from the particle. Furthermore, our potential $\Phi$ will be a solution to the adjoint Teukolsky equation with a source  \eqref{eq:OJ} that will be supported along the string. Our construction thereby gives a rather transparent conceptual understanding for the occurrence of such Dirac-type strings that have been observed in other approaches. It is also conceivable that our construction, which avoids the consideration of an interface, might provide a viable alternative scheme for numerical integrations. 

\subsection{Nonlinear perturbations: Decay and turbulent behavior}

By iterating our prescription, we can recursively obtain the $n$-th
order gravitational perturbation up to gauge as
$h^{(n)}_{ab} = g^{(n)}_{ab} + x_{ab}^{(n)} + \Re (\S^\dagger
\Phi^{(n)})_{ab}$ in terms of an $n$-order potential subject to a
sourced adjoint Teukolsky equation
$\O^{\dagger}\Phi^{(n)} = \eta^{(n)}$, where $\eta^{(n)}$,
$x_{ab}^{(n)}$ are determined by transport equations (ODEs) from
the nonlinear terms $T^{(n)}_{ab}$ in the $n$-th order Einstein tensor
(involving $h^{(1)}_{ab}, \dots, h^{(n-1)}_{ab}$), and where
$g^{(n)}_{ab}$ is an $n$-th order perturbation to another Kerr black
hole. While a thorough discussion is outside the scope of the present
paper, we here want to ask whether higher order perturbations could be
physically relevant in astrophysical situations.

At the linear level, an important aspect of characterizing black hole
perturbations is the determination of a quantized set of complex
frequencies associated with certain mode solutions to Teukolsky's
equation called quasinormal modes \cite{Nollert:1999ji}.  Quasinormal
modes are the natural resonances of black hole spacetimes and carry
information about the parameters of binary merger remnants.  Rapidly
spinning Kerr black holes exhibit a special family of weakly damped
quasinormal frequencies, which in a sense ``corotate'' with the
horizon. These modes each have real part proportional to the horizon
frequency and become arbitrarily long-lived in the extremal limit
\cite{Hod:2008zz,Yang:2013uba}.  At linear order, the long-lived modes
all get coherently excited, and their collective behavior leads to
interesting physical effects.  Near the horizon, the superposition of
modes results in a transient growth of tidal fields as measured by
infalling observers.  Away from the horizon, the mode sum yields a
transient power-law tail, slowing the asymptotic decay
\cite{Yang:2013uba, Gralla:2016sxp}.  Furthermore, as they have
commensurate real part of frequency, these modes can be
\emph{nonlinearly} in a state of resonance and can readily transfer
energy amongst each other.  As such, the nonlinear coupling between
multiple modes is conjectured to lead to turbulent cascades
exhibiting a Kolmogorov-type scaling~\cite{Yang:2014tla}. In this way,
such nonlinear features could in principle lead to a unique signature
in the ringdown signals of highly spinning black holes, as measured by
future gravitational wave detectors.

Generically, however, an arbitrary perturbation to Kerr will extract
angular momentum through the mechanism of superradiance, the wave
analog of the Penrose process, see e.g., \cite{Wald1984}. For
nonlinear mode-coupling effects to become important, it is, therefore,
necessary that they develop faster than (a) the time scale for linear
decay of the quasinormal modes, and (b) the spin-down time scale of
the black hole. We now give a heuristic argument, concluding that
as long as the perturbation amplitude and surface gravity are
appropriately chosen, both of these conditions can be satisfied.

We first estimate the spin-down time scale of the black
hole.\footnote{We thank S. Gralla for a discussion in which we made
  this estimate.} Suppose we have a mode perturbation to Kerr of
amplitude $A$, frequency $\omega$, and azimuthal number $m$. The
energy and angular momentum fluxes into the black hole scale
as~\cite{TeukolskyPress1974}
\begin{equation}
  \dot{M} \sim M^2 \omega(\omega-m\Omega_H) A^2, \qquad \dot{J} \sim M^2 m (\omega-m\Omega_H) A^2,
\end{equation}
where the overdot denotes differentiation with respect to ingoing
Kerr time and $\Omega_H$ is the angular velocity of the outer horizon.
For short,
\begin{equation}
\dot{M} \sim \omega_f \omega_n A^2, \qquad \dot{J} \sim M m \omega_n A^2 ,
\end{equation}
with dimensionless frequencies
\begin{equation}
\omega_f = M \omega, \qquad \omega_n=M(\omega-m\Omega_H).
\end{equation}
The fractional change in the extremality parameter (surface gravity)
$\kappa=\sqrt{1-J^2/M^4}$ due to absorption at the horizon is given by
\begin{IEEEeqnarray}{rCl}
  M\frac{\dot{\kappa}}{\kappa} &=& \frac{2J^2}{M^2\kappa^2}\left(\frac{\dot{M}}{M^2}-\frac{\dot J}{2MJ}\right)\nonumber\\
  &=& \frac{2J^2}{M^2\kappa^2}(\omega-m\Omega_H)A^2\left(\omega-\frac{M}{2J}m\right)\nonumber\\
  &=& \frac{2J^2}{M^2\kappa^2}(\omega-m\Omega_H)A^2\left(\omega-m\Omega_H + O(\kappa)\right).
\end{IEEEeqnarray}
In the last line, we have used that $\Omega_H=M/(2J)+O(\kappa)$. The
long-lived ``near-horizon'' modes of nearly extremal Kerr have
$\omega-m\Omega_H = O(\kappa)$, so this term does not change the order
of the final term. Therefore, for near-horizon modes, we have
\begin{align}
  M \frac{\dot{\kappa}}{\kappa} \sim \frac{A^2}{\kappa^2} \omega_n^2.
\end{align}
Thus $A\omega_n \ll \kappa$ suffices to make the fractional change in
$\kappa$ small over a typical ``fast'' time scale $M$. For
near-horizon modes, with $\omega_n\sim\kappa$, the spin-down time
scale goes as $\tau_{\text{spin-down}} \sim M/A^2$.

The time scale for linear decay of a near-horizon mode, meanwhile, is
$\tau_{\text{decay}}\sim M/\kappa$. Thus, for nonlinear mode-coupling
to become important, the perturbation amplitude must be such that the
mode-coupling time scale
$\tau_{\text{mode-coupling}} \lesssim \tau_{\text{spin-down}},
\tau_{\text{decay}}$.

The mode-coupling time scale depends on whether the fundamental
interaction involves a 3- or 4-mode
coupling~\cite{ZakharovBook}. Three-mode interactions are stronger,
but their presence depends on the dispersion relation of the modes and
any associated selection rules, analysis of which we do not attempt
here. Four-mode interactions are more generic. Three-mode interactions
have an interaction time scale that goes as $1/A$, whereas four-mode
interactions go as $1/A^2$. The overall strength of the interaction
depends on the precise overlap integrals, which would have to be
studied using the framework developed in~\cite{Green:2019a}. For now,
we assume this goes as $1/M$. By comparing these time scales to those
for dissipation and spin-down, we obtain constraints on the size of
the perturbation compared to the spin of the black hole in order for
nonlinear mode-coupling effects to become important (see
table~\ref{tab:mode cases}).
\begin{center}
\begin{table}
\begin{tabular}{| c | c | c| }
\hline
  & 3-mode $\tau\sim M/A$ & 4-mode $\tau\sim M/A^2$  \\ 
  \hline
  spin-down $\tau \sim M/A^2$ & $A\ll 1$ & --- \\  
 decay $\tau \sim 1/\kappa$ & $\kappa  \ll A$  & $\kappa \ll A^2$   \\ 
 \hline
\end{tabular}
\caption{\label{tab:mode cases}Criteria for long-lived quasinormal mode coupling.
  Conditions are given for three and four mode interactions such that
  the mode coupling time scales are longer than the linear decay and
  spin down time scales.}
\end{table}
\end{center}

To summarize this discussion, we conclude that if the leading
interaction involves three modes, then provided $1 \gg A \gg \kappa$,
nonlinear coupling could become important for near-extremal black
holes. If the leading interaction involves four modes and
$A^2 \gg \kappa$, then the interaction time scale is the same as the
spin-down time scale, and such nonlinear effects are possible,
although only marginally so. We will analyze this in detail in a
future paper \cite{Green:2019b}.

\subsection{Perturbative quantum gravity}

Our method should also bring substantial simplifications to
perturbative quantum gravity on the Kerr spacetime. For linear
perturbations, the corrector tensor $x_{ab}$ is zero, and our results
give a decomposition
$h_{ab} = \dot g_{ab} + \Re(\S^\dagger \Phi)_{ab}$ up to
gauge. It turns out that, with respect to the canonical symplectic
form $W$ for general relativity [cf.~\eqref{eq:W}], the zero mode
perturbations $\dot g_{ab}$ (finitely many degrees of freedom) are
symplectically orthogonal to the Hertz-potential
perturbations\footnote{This follows, e.g.,  from proposition 3.1 of
  \cite{Prabhu:2018jvy} and the fact that $\T \dot g = 0$.}
$(\S^\dagger \Phi)_{ab}$, and thus can be ``quantized''
independently. Thus, up to the zero mode perturbations (quantization
of a finite-dimensional phase space), the quantization of the
gravitational field reduces at linearized order to the quantization of
the Hertz-potential field. For these, we have the naturally induced
symplectic form
$\tilde W[\Phi_1, \Phi_2] = W[{\rm Re} \S^\dagger \Phi_1, {\rm Re} \S^\dagger
\Phi_2]$, which may be used as the basis for the quantization of the
field $\Phi$.

For the higher perturbative orders, we should be able to proceed in a recursive manner just as one does when defining the higher orders 
of the interacting Klein-Gordon quantum field in $\phi^4$-theory on a curved spacetime, say, see \cite{Hollands:2014eia,Hollands:2007zg}. The only difference should be that the sources of the equations for the $n$-th order perturbation are determined by a certain transport equation as described in section \ref{sec:5}, and that in addition to the Hertz-potential quantum field $\Phi$, we need to keep track of the corrector $x_{ab}$ at each perturbation order, also as described in section \ref{sec:5}. We leave the details to a future investigation. 

\medskip

This paper is organized as follows. In section \ref{sec:2} we review the essential parts of the GHP formalism for the convenience of the reader. 
In section \ref{sec:3} we study the decomposition \eqref{eq:decintro} for solutions to the homogeneous linearized Einstein equations. In section \ref{sec:4}, we generalize this to solutions to the linearized Einstein equations with source. In section \ref{sec:5} we point out the fairly obvious 
application to higher order (nonlinear) perturbations. Some formulae are relegated to the appendix. 

{\bf Conventions:} We use the $(+---)$ signature convention for the metric and the conventions of \cite{Wald1984} otherwise. We also make extensive use of the Geroch-Held-Penrose (GHP)-formalism \cite{Geroch:1973am} and the associated standard notation. 

\section{GHP formalism and Teukolsky equation}
\label{sec:2}

Our construction makes heavy use of the special geometric features of the background Kerr geometry. The key simplifying feature behind most arguments is in fact the 
algebraically special nature of the background geometry and, of course, the vacuum Einstein equation
\ben
R_{ab} = 0,
\een
assumed throughout this article.
We recall (see \cite{Wald1984} for details) that the Weyl tensor $C_{abcd}$ of a 4-dimensional Lorentz metric generically has four distinct ``principal null directions'' 
at each point, where a principal null direction is a null vector $k^a$ such that 
\ben
0=k_{[f}C_{a]bc[d} k_{e]} k^b k^c.  
\een
\begin{definition}
A 4-dimensional  spacetime is called algebraically special (or Petrov type II) if two principal null directions coincide at every point, $k_1^a = k^a_2 = l^a$, and in this case $C_{abc[d} l_{e]} l^b l^c=0$. An algebraically special spacetime is called Petrov type D if there is another pair of coinciding principal null directions $k_3^a = k^a_4 = n^a$. 
\end{definition}

In an algebraically special spacetime, it is true that $l^a$ is tangent to a congruence of shear free null-geodesics by the Goldberg-Sachs theorem \cite{GoldbergSachs2009}. Examples of 
such spacetimes are the Kerr- and Robinson-Trautman families of metrics \cite{RobinsonTrautman1962}. The former ones are actually of type D, while in the latter the twist of $l^a$ vanishes. In order to take full advantage of the algebraically special property, it is useful to consider null tetrads $(l^a, n^a, m^a, \bar m^b)$ such that $l^a$ is aligned with the principal null direction (and 
$n^a$ with the other one in type D). Our conventions for the tetrad are: $l^a n_a = 1$, $m^a \bar m_a = -1$ and all other inner product vanish. From the null tetrad, we define as usual the Weyl components,
\ben
\begin{split}
\label{eq:Weylcomp}
\Psi_0 & = -C_{abcd} \, l^a m^b l^c m^d, \\
\Psi_1 & = - C_{abcd} l^a n^b l^c m^d, \\
\Psi_2 & = -\half C_{abcd}( l^a n^b l^c n^d + l^a n^bm^c \mb^d),\\
\Psi_3 & = -C_{abcd} l^a n^b \mb^c n^d, \\
\Psi_4 & = -C_{abcd} \, n^a \bar m^b n^c \bar m^d,
\end{split} 
\een
and the spin coefficients $\rho, \rho', \sigma, \sigma', \tau, \tau', \kappa, \kappa', \beta, \beta', \epsilon, \epsilon'$; see appendix \ref{sec:B}. 
The real part of $\rho$ corresponds to the expansion of the null-geodesic congruence tangent to $l^a$, the imaginary part to the twist, and the real and imaginary parts of $\sigma$ correspond to the shear of $l^a$. $\kappa$ measures the failure of the congruence to be geodesic in general, so by the Goldberg-Sachs theorem \cite{GoldbergSachs2009} and its corollaries,
\ben
\label{eq:GSthm}
\kappa = \sigma = \Psi_0 = \Psi_1 = 0. 
\een 
The corresponding primed quantities correspond to sending $l^a \to n^a$ and $m^a \to \bar m^a$. By choosing $n^a$ appropriately, one can set further quantities to zero, but there are different options:
\begin{itemize}
\item[n1)] In any type II spacetime, we can perform a null rotation $l^a \to l^a, m^a \to m^a + A l^a, n^a \to A\bar m^a + \bar A m^a + A\bar A l^a$, under which $\tau$ transforms as 
$\tau \to \tau + A\rho$. Then, if $\rho \neq 0$, we can adjust $A$ so that $\tau = 0$, and this choice fixes the null vector $n^a$ up to a rescaling. Making further use of the Einstein field equation, one then sees that $\tau' = \sigma'=0$  in addition to \eqref{eq:GSthm}; see \cite{Held:1975} for details. 

\item[n2)] In type D, we may alternatively choose $n^a$ to be the second principal null direction, which results in the primed version of the Goldberg-Sachs theorem 
$\kappa' = \sigma' = \Psi_3 = \Psi_4 = 0$ in addition to \eqref{eq:GSthm}. In Kerr, an example of such a tetrad is the Kinnersley frame; see 
appendix \ref{sec:B}.
\end{itemize}

Except for the case of Schwarzschild spacetime, one must in general choose between these two possibilities, and we shall make use of either one of them in the following depending on the purpose. 

The GHP formalism \cite{Geroch:1973am} allows one to handle in a very efficient manner the equations obtained for the spin coefficients and Weyl components. Furthermore, that formulation has a geometrical basis which renders its equations automatically invariant under the remaining permissible rescalings of the tetrad $(l^a, n^a, m^a, \bar m^b)$. 
These rescalings consist of (a) a local boost, sending $l^a \to \lambda \bar \lambda l^a$, $n^a \to \lambda^{-1} \bar \lambda^{-1} n^a$, 
or (b) a local rotation $m^a \to \lambda \bar \lambda^{-1} m^a$, where $\lambda$ is a nonzero complex number depending on the spacetime point. 
A GHP quantity $\eta$ is said to have weights $(p,q)$ if transforms 
as 
\ben 
\eta \to \lambda^p \bar \lambda^q \eta \quad \Longleftrightarrow: \quad \eta \GHPwt (p,q),
\een
under a combined local boost and rotation. 

More mathematically speaking, the construction can be seen as
follows. Given two null {\em directions}---rather than
vectors---aligned with the given $l^a, n^a$ at each point, one can
define the bundle of {\em all} null frames
$(l^a, n^a, m^a, \bar m^b)$ over $\sM$ such that $l^a, n^a$ point
along the prescribed directions and such that $m^a, \bar m^a$ are
positively oriented and span the complexified orthogonal complement of
the plane determined by $l^a, n^a$. This set is seen to define a
principal fibre bundle, $\sP$, over $\sM$ with structure (gauge) group
the boosts and rotations, i.e., $G=\RR_+ \times SO(2)$, isomorphic also
to the multiplicative group $G=\CC_\times$ of nonzero complex
numbers; see \cite{kobayashi_nomizu} for the basic definitions related
to the notion of a principal fibre bundle. The gauge group $G$ acts
precisely by local boosts and rotations on $\sP$. $G=\CC_\times$ also
acts by multiplication with $\lambda^p \bar \lambda^q$ on the
1-dimensional vector space $\CC$, and this gives a representation
$\pi_{p,q}$. The GHP quantities $\eta \GHPwt (p,q)$ are precisely the
sections of the so-called associated complex line bundle
$\sL^{p,q} = \sP \rtimes_{\pi_{p,q}} \CC$. We may also consider
``mixed'' tensorial/GHP objects of GHP weight $(p,q)$ and tensorial
rank $(r,s)$, which are sections in the vector bundles
$\sL^{p,q} \otimes T^{r,s} \sM$.

The key idea of the GHP formalism is to view the spin coefficients
$\rho, \rho', \kappa, \kappa', \tau, \tau', \sigma, \sigma'$ as GHP
quantities of the appropriate weight as specified in appendix
\ref{sec:B}. Likewise, the tetrad components of an ordinary tensor are
GHP quantities of an appropriate weight. For instance, if $\xi_a$ is a
covector field then $\xi_l = \xi_a l^a \GHPwt (1, 1)$, or
$\xi_{n} = \xi_a n^a \GHPwt (-1, -1)$, or $\rho \GHPwt (1,1)$, etc.
The remaining spin coefficients $\beta, \beta', \epsilon, \epsilon'$
by contrast are used to define a covariant derivative operator (i.e., a
connection) called $\Theta_a$ of the vector bundle
$\sL^{p,q} \otimes T^{r,s} \sM$, where
$T^{r,s} \sM = T\sM^{\otimes r} \otimes T^*\sM^{\otimes r}$. Its
definition is
\begin{align}
  \label{nabladef}
  \Theta_a
  &= \nabla_a - \half(p+q) n^b \nabla_a l_b + \half(p-q) \mb^b \nabla_a m_b \nonumber \\
  &=\nabla_a + l_a(p\epsilon'+q\bar \epsilon') + n_a(-p\epsilon-q\bar \epsilon)
    - m_a(p \beta' -q \bar \beta) - \bar m_a(-p \beta +q \bar \beta'). 
\end{align}
Ordinary tensor fields are identified with sections of $\sL^{0,0} \otimes T^{r,s} \sM$, i.e., mixed GHP-tensor fields of weight $(0,0)$.  
The GHP-covariant directional derivatives along the tetrad legs are denoted traditionally by
\ben
\th = \Theta_l, \quad 
\th' = \Theta_n, \quad 
\edth = \Theta_m, \quad 
\edth' = \Theta_{\bar m} . 
\een
These operators shift the GHP weights by the amounts $\th: (1,1), \th': (-1,-1), \edth: (1,-1), \edth': (-1,1)$.
Apart from mathematical elegance, this geometric 
 viewpoint is useful because we may encounter null tetrads that behave in a singular way at certain points of $\sM$, such as the horizon, infinity, the north pole of a sphere, etc. 
The components of a mixed GHP quantity in such a singular null frame will therefore also be singular, but such a singularity is obviously absent in the geometrical viewpoint where it is simply ascribed to a bad choice of gauge. By working throughout with invariantly-defined GHP quantities, the formalism therefore automatically takes care of such artificial singularities. 

As noted by Wald \cite{Wald:1978vm}, the essence of the Teukolsky
formalism \cite{TeukolskyLett1972,Teukolsky:1973ha} and its extension
by \cite{Chrzanowski:1975wv,Kegeles:1979an} may be succinctly encoded
in an operator relation. Let $(\sM, g_{ab})$ be a type II vacuum
solution.  To state the operator relation, we introduce the linearized
Einstein operator,
\begin{equation}\label{eq:E}
  (\E h)_{ab} =  
  \half \left[ -\nabla^c \nabla_c h_{ab} - \nabla_a \nabla_b h^c_{\phantom{c}c} + 2 \nabla^c \nabla_{(a}h_{b)c} + g_{ab} \left( \nabla^c \nabla_c h^d_{\phantom{d}d} - \nabla^c \nabla^d h_{cd}\right) \right],
\end{equation}
as well as partial differential
operators $\S, \T, \O$ defined as follows. $\S, \T$ act on symmetric
rank-two covariant tensor fields and give a GHP scalar of weight
$(4,0)$,
\begin{subequations}
  \begin{align}\label{eq:S}
    \S T ={}& (\edth - \bar \tau' - 4\tau)\left[(\th - 2\bar \rho) T_{lm} - (\edth - \bar \tau') T_{ll} \right] \nonumber \\
            & + (\th - \bar \rho - 4\rho)\left[(\edth - 2\bar \tau') T_{lm} - (\th - \bar \rho) T_{mm} \right], \\
    \label{eq:T}\T h ={}& \half(\edth-\bar \tau')(\edth-\bar \tau') h_{ll} + \half (\th - \bar \rho)(\th - \bar \rho) h_{mm} \nonumber \\
            &-\half\left[(\th- \bar \rho)(\edth-2\bar \tau') + (\edth -\bar \tau')(\th -2\bar \rho)\right] h_{(lm)}.
  \end{align}
\end{subequations}
$\O$, the wave operator appearing in Teukolsky's master equation \cite{Teukolsky:1973ha, TeukolskyLett1972}, takes a $(4,0)$ GHP scalar to another $(4,0)$ GHP scalar and is defined by
\begin{equation}\label{eq:O}
  \O \eta = 2 \left[ (\th - 4\rho - \bar \rho)(\th'-\rho') - (\edth - 4\tau - \bar \tau')(\edth' - \tau') -3\Psi_2 \right] \eta,
 \end{equation}
The operator relation is
\ben
\label{eq:OT}
\O \T = \S \E , 
\een   
and can be exploited as follows. The quantity $\T h=\psi_0$ is by construction equal to the linearization of the Weyl component $\Psi_0$. (We denote the perturbed values of the Weyl scalars $\Psi_n$ given in \eqref{eq:Weylcomp} as $\psi_n$.) 
Then, if $h_{ab}$ is a solution 
to the linearized field equation $\E h_{ab} = 0$, it follows that $\psi_0$ solves the homogeneous Teukolsky equation $\O \psi_0 = 0$.  We can also take the formal adjoint of this operator relation and use that $\E^\dagger = \E$. [The formal adjoint\footnote{By contrast with the conventions in quantum mechanics, the operation $\dagger$
is linear, rather than anti-linear. This is because no sesqui-linear inner product is used in the definition.} 
$\P^\dagger$ of a partial differential operator $\P$ taking sections in $\sL^{p,q} \otimes T^{r,s} \sM$ 
to sections in $\sL^{p',q'} \otimes T^{r',s'} \sM$ is a partial differential operator taking sections in $\sL^{-p',-q'} \otimes T^{s',r'} \sM$
to sections in $\sL^{-p,-q} \otimes T^{s,r} \sM$. It is defined uniquely by the condition that
\ben
(\P^\dagger \eta')^{a \dots b}{}_{c \dots d} \eta_{a \dots b}{}^{c \dots d} - \eta'^{a \dots b}{}_{c \dots d} (\P \eta)_{a \dots b}{}^{c \dots d} = \nabla_a w^a
\een
for some vector field $w^a$ constructed locally out of $\eta, \eta'$ and their derivatives of GHP weight zero.] This results in the relation
\ben
\label{eq:OTdag}
\T^\dagger \O^\dagger = \E \S^\dagger. 
\een
As a consequence, suppose $\Phi \GHPwt (-4,0)$ satisfies $\O^\dagger \Phi= 0$. Then $h_{ab} = (\S^\dagger \Phi)_{ab}$ is a complex-valued solution to the linearized Einstein equation, $\E h_{ab} = 0$. By taking the real part and using that $\E$ is a real operator, we thereby get a real tensor field satisfying the linearized Einstein equation from the ``Hertz potential'' $\Phi$. The adjoint Teukolsky operator $\O^\dagger$ is explicitly
\begin{align}
  \label{eq:Odag}
  \O^\dagger \Phi ={}&  2\left[ (\th'-\rho')(\th + 3 \rho) - (\edth' - \tau')(\edth +3\tau) -3\Psi_2 \right] \Phi \nonumber \\
  ={}& 2\left[ \left(\frac{(r^2+a^2)^2}{\Delta}-a^2 \sin^2\theta \right) \frac{\pd^2 \Phi}{\partial t^2} +\frac{4 M ar }{\Delta } \frac{\pd^2 \Phi}{\pd t \pd \phi}  +4 \left( \frac{M(r^2 -a^2)}{\Delta} -r -ia \cos\theta \right) \frac{\pd \Phi}{\pd t} \right. \nonumber\\
                     &+ \left( \frac{a^2}{\Delta} - \frac{1}{\sin^2\theta} \right) \frac{\pd^2 \Phi }{\pd \phi^2} - \Delta^{2} \frac{\pd}{\pd r}\left( \Delta^{-1} \frac{\pd \Phi}{\pd r} \right) - \frac{1}{\sin \theta} \frac{\pd}{\pd \theta} \left( \sin \theta \frac{\pd \Phi}{\pd \theta} \right) \nonumber \\
                     & \left. + 4 \left( \frac{a(r-M)}{\Delta} + \frac{i \cos \theta}{\sin^2 \theta} \right) \frac{\pd \Phi}{\pd \phi} + \left(4 \cot^2\theta +2\right) \Phi \right],
\end{align}
where the second line is the coordinate expression in the Kinnersley tetrad in the Kerr spacetime, see appendix \ref{sec:B}. The adjoint operators $\T^\dagger, \S^\dagger$ are explicitly
\begin{align}\label{eq:Tdag}
  (\T^\dagger \eta)_{ab} ={}
  & \half l_a l_b(\edth-\tau)(\edth-\tau) \eta + \half m_a m_b (\th-\rho)(\th-\rho) \eta \nonumber\\
  & - \half l_{(a} m_{b)}\{(\edth + \tab' -\tau)(\th - \rho) + (\th - \rho + \bar \rho)(\edth-\tau) \} \eta 
\end{align}
and 
\begin{align}\label{eq:Sdag}
  (\S^\dagger \Phi)_{ab} ={}
  & - l_a l_b(\edth-\tau)(\edth+3\tau) \Phi - m_a m_b (\th - \rho)(\th + 3\rho) \Phi \nonumber \\
  & + l_{(a} m_{b)}\{(\th - \rho + \bar \rho)(\edth + 3\tau) + (\edth-\tau+\bar \tau')(\th +3\rho) \} \Phi . 
\end{align}
As one sees from this expression, the real perturbation $h_{ab} = \Re(\S^\dagger \Phi)_{ab}$ automatically is in a gauge where 
\ben
h_{ab} l^b = 0 = g^{ab} h_{ab} \qquad (= h_{ab} m^a \bar m^b). 
\een 
We follow the tradition in which this gauge is referred to as ``ingoing radiation gauge'' (IRG), although in a general type II vacuum spacetime this terminology has little physical meaning.

\section{Decomposition of $h_{ab}$ for homogeneous equation}
\label{sec:3}

Although we are generally interested in the inhomogeneous linearized
Einstein equation~\eqref{eq:LEinstein}, we will in this section first
inspect the homogeneous equation
\begin{equation}
  (\E h)_{ab}  = 0 . 
\end{equation}
As we have
recalled, on a type II vacuum background, if $\Phi$ is a (IRG) Hertz
potential, i.e., a weight $(-4,0)$ GHP scalar satisfying the (adjoint)
homogeneous Teukolsky equation $\O^\dagger \Phi = 0$, then the metric
perturbation $h_{ab} = \Re(\S^\dagger \Phi)_{ab}$ is a solution to the
source-free linearized Einstein equation. If we add to such a
solution a pure gauge perturbation
$(\Lie_\xi g)_{ab} = \nabla_a \xi_b + \nabla_b \xi_a$ and a zero mode
perturbation $\dot g_{ab} = \frac{\dd}{\dd s}g^{M(s),a(s)}_{ab}$, we
obtain a new solution,
\begin{equation}
  \label{eq:decomp}
  h_{ab} = \Re(\S^\dagger \Phi)_{ab} + (\Lie_\xi g)_{ab} + \dot g_{ab},
\end{equation}
and we would now like to argue that, in a sense, this is the most
general solution in the case of the Kerr background, or more
precisely, when $(\sM, g_{ab})$ is the exterior region of the black
hole in the Kerr spacetime.

We shall give an argument for \eqref{eq:decomp} following that given by Prabhu and Wald \cite{Prabhu:2018jvy} in the case of Schwarzschild, with suitable modifications for Kerr. Their argument is of a conceptual nature as it is embedded in the phase-space (Hamiltonian) formulation of general relativity. In its basic form, it applies to the 
sub-extremal family of Kerr black holes $|a|<M$, although later we will need to impose, for technical reasons, the more stringent condition $|a|\ll M$, which should be unnecessary.
One begins by defining the symplectic current for vacuum general relativity. Let $h_{ab}^{}$, $h_{ab}'$ be two symmetric tensor fields, not necessarily solutions to the linearized Einstein equations. By the definition of the adjoint, since $\E^\dagger = \E$, we have 
\ben
(\E h')^{ab} h_{ab} - (\E h)^{ab} h_{ab}' = \nabla_a w^a, 
\een
where $w^a$ is in the present context called the ``symplectic current.'' Evidently, it is conserved if $h_{ab}^{}, h_{ab}'$ solve the linearized Einstein equation. The explicit form 
of $w^a$ can be found e.g., in \cite{Iyer:1994ys}. Following that reference, one defines the symplectic form of general relativity to be 
\ben
\label{eq:W}
W[h, h'] = \int_\Sigma \nu^a w_a(h, h') ,  
\een
where $\Sigma$ is a Cauchy surface with unit future pointing normal $\nu^a$, stretching from the bifurcation surface $\sS = \sH^- \cap \sH^+$ to spatial infinity. Since $w^a$ is conserved, the definition of $W$ is invariant under deformations of $\Sigma$ leaving $\sS$ and spatial infinity fixed. If $\xi^a$ is a smooth vector field vanishing at $\sS$
and approaching as $r\to \infty$ either a time translation, spatial translation, or rotation [in the asymptotically Cartesian coordinate system $(t, x_1, x_2, x_3)$ determined by $(t,r,\theta,\phi)$], the symplectic inner products with the corresponding pure gauge perturbation $(\Lie_\xi g)_{ab}$,
\ben
W[h, \Lie_\xi g] = 
\begin{cases}
\dot M & \text{if $\xi^a \to \partial_t$}\\
-\dot P_i & \text{if $\xi^a \to \partial_i$}\\
\dot J_i & \text{if $\xi^a \to \epsilon_{ijk} x_k\partial_j$}, 
\end{cases}
\een
define the perturbed ADM mass, linear momentum, respectively angular momentum, see \cite{Iyer:1994ys}. 

From a solution to the Einstein equation we may determine its initial data $(q_{ab}, p^{ab})$, where $q_{ab}$ is a smooth symmetric tensor field on $\Sigma$ representing the  intrinsic metric, and where $p^{ab}$ is a smooth symmetric tensor density on $\Sigma$ representing the conjugate momentum (related to the perturbed extrinsic curvature), 
see \cite{Wald1984}. The corresponding quantities for the linearized perturbation are denoted by $(\dot q_{ab}, \dot p^{ab})$, where here and in the following, 
an overdot refers to a first order perturbed quantity, and not a time derivative.  In terms of these, the symplectic form reads
\ben
W[\dot q,\dot p ; \dot q',\dot p'] = \int_\Sigma \dot p^{ab} \dot q_{ab}' - \dot p^{ab'} \dot q_{ab} . 
\een
The symplectic form is automatically finite on the space of square integrable pairs $(\dot q_{ab}, \dot p^{ab})$. We denote the corresponding $L^2$-type norm by $\| \ . \ \|$, 
so 
\ben
\| (\dot p, \dot q)\|^2 = \int_\Sigma \dot p^{ab} \dot p_{ab} (-q)^{-1/2} + \dot q^{ab} \dot q_{ab} (-q)^{+1/2} .
\een
If $\sX$ is a linear subspace of $L^2 \times L^2$ of such pairs, we denote by $\overline{\sX}$ its closure in the norm $\| \ . \ \|$. 

Following section 4.1 of \cite{Hollands:2012sf} we consider several such linear subspaces in this $L^2 \times L^2$ space
of (unconstrained) initial data. We let $\sW$  be the space of all gauge perturbations $(\Lie_\xi g)_{ab}$ (identified with initial data)
generated by smooth vector fields $\xi^a$ that become an asymptotic
translation or rotation at infinity and whose projection onto the bifurcation surface $\sS$
vanishes. Further, we let $\sV = (\sW)^\perp$, where $\perp$ denotes the symplectic orthogonal complement. 
Finally, we can introduce within $\sV$ the subspace $\sV \cap \sU$, where $\sU \subset C^\infty \times C^\infty$ is 
a certain intersection of weighted Sobolev spaces. It is shown in section 4.1 of \cite{Hollands:2012sf} that: (i)
Initial data from $\sV \cap \sU$ satisfy the linearized constraints, (ii) have
$\dot M = \dot P = \dot J = 0$, (iii) $\sU \cap \sV$ is dense in the closed subspace $\sV$, (iv) satisfy
$\dot q_{ab}, \dot p^{ab} = o(r^{-3/2})$ in an asymptotically Cartesian coordinates system as $r \to \infty$, 
with derivatives falling off faster by corresponding powers of $r^{-1}$ and (v)
\ben
\label{bndy}
\dot \varepsilon = \dot \vartheta_l = \dot \vartheta_n = 0, \quad \text{on $\sS$}
\een
where $\varepsilon$ is the induced area element on the bifurcation surface $\sS$ and $\dot \vartheta_{n,l}$ the perturbed expansions
in the directions $n^a, l^a$ respectively. The second two conditions physically mean that the perturbation is in a gauge 
in which the horizons $\sH^\pm$ stay in a fixed position, whereas the first is just a gauge condition that can be imposed 
consistent with $\dot M = \dot J = 0$ in view of the first law of black hole mechanics, see \cite{Hollands:2012sf} for further explanations.

Finally, let
$\sY$ denote the space of perturbations $  h_{ab} = \Re(\S^\dagger \Phi)_{ab}$ generated by smooth GHP scalar fields $\Phi$ of type
$(-4,0)$ such that $\O^\dagger \Phi=0$ and such
that $\Phi$ has compact initial data on $\Sigma$. Prabhu and Wald
\cite{Prabhu:2018jvy} now argue as follows. Suppose $h_{ab}$ is 
a smooth solution to the linearized Einstein equations with initial data 
$(\dot q_{ab}, \dot p^{ab}) \in \sV \cap \sU$, and suppose one could show that the relation
\begin{equation}
  \label{eq:symplectic_orth}
  W[h, \Re \S^\dagger \Phi] = 0 \qquad \forall \Re(\S^\dagger \Phi)_{ab} \in \sY
\end{equation}
implies that $h_{ab}$ is pure gauge perturbation in $\sW$. In formulas, this means
\ben
\label{eq:concl}
(\sY)^\perp \cap (\sU \cap \sV) \subset \sW.
\een
Now take the symplectic complement $\perp$ of this relation. We can use $(\sU \cap \sV)^\perp = \sV^\perp$ in view of (iii), 
and further that $\sV^\perp = \sW^{\perp\perp} = \overline \sW$, by definition. Then we obtain 
$\overline{(\sY + \sW)} \supset \sV$. Thus, in view of (i)--(v), we would learn that:

\medskip
\noindent
{\bf Conclusion:} \cite{Prabhu:2018jvy} {\em Any smooth perturbation 
with $\dot M = \dot P = \dot J = 0$, with \eqref{bndy} and the falloff for $r \to \infty$ implied by the Sobolev norm 
of $\sU$ (basically $\dot q_{ab}, \dot p^{ab} = o(r^{-3/2})$) can be approximated in the norm $L^2$ on initial data
by the sum of a Hertz potential perturbation  $\Re(\S^\dagger \Phi)_{ab}$ and a pure gauge perturbation ${\mathcal L}_\xi g_{ab}$.
Since
perturbations with a given nonzero $\dot M, \dot P, \dot J$ can be
obtained by adding a perturbation $\dot g_{ab}$ towards another,
perhaps boosted/rotated, Kerr black hole this gives an argument---and
precise sense---in which \eqref{eq:decomp} holds true.} 

\medskip
\noindent
Furthermore, \cite{Prabhu:2018jvy} have given 
arguments that \eqref{eq:concl} must in fact hold
in the Schwarzschild background.  Thus, in the Schwarzschild
spacetime, this completes the argument.
We now give a corresponding argument for the Kerr metrics with sufficiently small $|a| \ll  M$.  First, 
it has been demonstrated in \cite{Hollands:2012sf} by gluing methods that $\sV \cap \sU$ has the same closure as those constrained initial data in 
$\sV$ that are smooth and vanish outside some value $r>r_0$ on $\Sigma$. We may therefore restrict to such solutions, and call the corresponding space of
initial data $\sV_0$. Secondly, \cite{Prabhu:2018jvy} have shown that in a general type II background, the relation \eqref{eq:symplectic_orth} implies $\T h=0$, so $\psi_0=0$. The argument is then completed by the following theorem, which shows that $h_{ab}$ must be pure gauge.

\begin{theorem}
  Let $h_{ab}$ be a solution to the linearized Einstein equation
  on a Kerr background with $|a| \ll  M$, with initial data
  $(\dot q_{ab}, \dot p^{ab}) \in \sV_0$, having $\psi_0=0$. Then also
  $\psi_4=0$ and $h_{ab}=\Lie_\xi g_{ab}$ is a pure gauge perturbation.
\end{theorem}

{\bf Remarks:} 1) The proof does not show explicitly that the gauge vector field must be in $\sW$, i.e. it need not have vanishing projection onto the 
bifurcation surface. This means that the space for which the approximation claimed in the Conclusion could be slightly smaller than $\sV$, but it will always contain
e.g. perturbations with compact support on the initial data surface $\Sigma$. 

2) Before we give the proof of this theorem, we point out that Wald \cite{Wald:1973jmp} has given an argument for smooth mode solutions. Our proof thus 
generalizes that of Wald in the sense that our solutions need not be mode solutions. 

3) The restriction to $|a| \ll M$ comes about because we are using
results on the large time decay of solutions to the Teukolsky equation
from~\cite{Dafermos:2017yrz,Ma:2017bxq}, which hold under
this condition. Concretely, we use that integrals of the type
$\int_{\{r_1<r<r_2, u_1<u<u_2\}} |\Psi^{-4/3}_2 \psi_4|^2 \sin\theta \dd u \dd r \dd \varphi_* \dd \theta$ 
are uniformly bounded in $u_2$ for fixed $u_1$ and $r_+ < r_1< r_2 < \infty$.
Our proof would generalize to all $|a|<M$ if such a 
result could be generalized to this range.\footnote{To be precise, this is not a GHP-invariant statement. We here mean the outgoing Kerr-Newman coordinates and e.g., the 
Kinnersley frame. An invariant statement would be $\int_{\{r_1<r<r_2, u_1<u<u_2\}} \rho^8 |\Psi^{-4/3}_2 \psi_4|^2 \, \dd vol_g < \infty$.}

\begin{proof}
  By the Teukolsky-Starobinsky identities which hold in a general type
  D background (see e.g.,
  \cite{PriceThesis},\cite{Aksteiner:2016pjt}),
  \begin{subequations}
    \begin{align}
      \label{eq:TeukStar}
      \th^4 \Psi_2^{-4/3} \psi_4 &= \edth^{\prime 4}  \Psi_2^{-4/3} \psi_0 - 3 \GHPLie_\xi \bar \psi_0, \\
      \th^{\prime 4} \Psi_2^{-4/3} \psi_0 &= \edth^{4}  \Psi_2^{-4/3} \psi_4 + 3 \GHPLie_\xi \bar \psi_4,
    \end{align}
  \end{subequations}
where $\xi^a = \Psi^{-1/3}_2(\tau' m^a - \tau \mb^a - \rho' l^a + \rho n^a)$ (which is $M^{-1/3} \partial_t$ in Kerr), and where $\GHPLie$ is the 
GHP-covariant Lie derivative~\cite{edgar2000integration}. Because $\psi_0=0$, these relations evidently give strong constraints on $\psi_4$. In the remainder of the proof 
we will argue that, in fact, we learn $\psi_4=0$.

From \eqref{eq:TeukStar}, we first conclude in view of $\th \rho = \rho^2$ that
\ben
\label{eq:poly}
 \Psi_2^{-4/3} \psi_4  = \alpha_0^\circ + \alpha_1^\circ \frac{1}{\rho} +\alpha_2^\circ  \frac{1}{\rho^2}  +\alpha_3^\circ  \frac{1}{\rho^3} .  
\een 
Here and in the following, a degree mark $\circ$ on a quantity indicates that this quantity is annihilated by $\th$, 
so $x^\circ$ means that  $\th x^\circ = 0$. For the solution $\Psi_2^{-4/3}\psi_4$ to the adjoint Teukolsky equation, we then use the following consequence of the 
Teukolsky-Starobinsky identities, see   \cite{PriceThesis},
\ben
\label{eq:TSI2}
\th^{\prime 4} \bar \Psi_2^{-4/3} \th^4 \Psi_2^{-4/3} \psi_4 =
\edth^{\prime 4} \bar \Psi_2^{-4/3}  \edth^4 \Psi_2^{-4/3} \psi_4-
9 \bar \GHPLie_\xi \GHPLie_\xi \psi_4 .
\een
Using \eqref{eq:poly}, the left side vanishes. Since $\GHPLie_\xi \Psi_2 = 0$ in Kerr, this gives
\ben
\label{eq:Algspe0}
0=\{\Psi_2^{-4/3} \edth^{\prime 4} \bar \Psi_2^{-4/3}  \edth^4 -
9\bar \GHPLie_\xi \GHPLie_\xi\} \left(  \alpha_0^\circ + \alpha_1^\circ \frac{1}{\rho} +\alpha_2^\circ  \frac{1}{\rho^2}  +\alpha_3^\circ  \frac{1}{\rho^3} \right).
\een
To take full advantage of $\th \alpha_i^\circ = 0$ in this equation and in several similar instances in the rest of the paper, we employ a formalism invented by 
Held \cite{Held:1974,Held:1975}. His formalism has three main ingredients. The first is a set of new operators, denoted by $\Dbar, \Dbar'$ and $\Nbar$ which replace $\edth, \edth'$ and $\th'$ and contrary to the latter have the property that $\Dbar \, x^\circ, \Dbar' x^\circ$ and $\Nbar x^\circ$ are quantities annihilated by $\th$. The second is a list of identities, generated from the Einstein equations and Bianchi identities, expressing any background quantity in terms of $\rho$ and quantities annihilated by $\th$, together with a table of how the new operators act on these. These two features enable one to write any expression appearing in \eqref{eq:Algspe0} as a Laurent polynomial in $\rho$ with coefficients that are annihilated by $\th$. 
The third ingredient is a simple lemma showing that if GHP quantities $a^\circ_i$ satisfy 
\ben
\label{eq:poly1}
a_0^\circ + a_1^\circ \rho + \dots +  a_n^\circ \rho^n = 0 \quad \Longrightarrow \quad a^\circ_i=0.
\een
[Proof: Divide by $\rho^n$, then act with $\th^n$ and use repeatedly $\th \rho = \rho^2$ to obtain $a^\circ_0=0$. Substitute this relation and 
iterate the process.]
We begin by recalling the precise definition of the new operators \cite{Held:1974} on weight $(p,q)$ GHP quantities,
\begin{subequations}
  \label{eq:Hops}
  \begin{align}
    \Nbar &=\th - \tab \Dbar - \tau \Dbar' + \tau \tab \left( \frac{q}{\bar \rho} + \frac{p}{\rho} \right) + \frac{1}{2} \left( \frac{q \bar \Psi_2}{\bar \rho} + \frac{p \Psi_2}{\rho} \right), \\
    \Dbar &=\frac{1}{\bar \rho} \edth + \frac{q\tau}{\rho}, \\
    \Dbar' &= \frac{1}{\rho} \edth' + \frac{p\tab}{\bar \rho}.
  \end{align}
\end{subequations}
Using the vacuum Einstein equations in GHP form, \cite{Held:1974} next shows that on a type D background of the Case II in  Kinnersley's classification~\cite{Kinnersley:1969zza} (which includes Kerr) with a null tetrad satisfying n2),
\begin{subequations}
  \label{eq:exp}
  \begin{align}
    \rho' &= \rho^{\prime \circ} \bar \rho - \half \Psi_2^\circ \rho^2 - (\Dbar \tab^\circ + \half \Psi_2^\circ) \rho \bar \rho - \tau^\circ \tab^\circ \rho^2 \bar \rho ,\\
    \tau' &= -\tab^\circ \rho^2, \\
    \tau &= \tau^\circ \rho\bar\rho, \\
    \Psi_2 &= \Psi_2^\circ \rho^3,
  \end{align}
\end{subequations}
with $\tau^\circ \GHPwt (-1,-3)$, $\Psi_2^\circ \GHPwt (-3,-3)$, $\rho^{\prime \circ} \GHPwt (-2,-2)$, together with 
\ben
\label{eq:barrho}
\frac{1}{\bar \rho} = \Omega^\circ + \frac{1}{\rho}, 
\een
with $\Omega^\circ \GHPwt (-1,-1)$.
These identities are complemented by identities for the action of the new operators on $\rho$ and on  $\Omega^\circ, \rho^{\prime \circ}, \tau^\circ, \Psi_2^\circ$ given in the table \ref{tab:1} extracted from \cite{Held:1974}. 

\begin{table}
\begin{tabular}{ | c | c | c | c |}
\hline
& $\Nbar$ & $\Dbar$ & $\Dbar'$ \\ \hline \hline
$\rho$ & 
$\rho^2 \rho^{\prime \circ} - \half \rho(\rho \Psi_2^\circ + \bar \rho \bar \Psi_2^\circ) - \rho^3 \bar \rho \tau^\circ \bar \tau^\circ$ &
$\tau^\circ \rho^2$ &
$-\bar \tau^\circ \rho^2$ \\
$\Omega^\circ$ & 0 & $2\tau^\circ$ & $-2\tab^\circ$\\ 
$\rho^{\prime \circ}$ & 0 & 0 & 0 \\
$\tau^\circ$ & 0 & 0 & 
                       $\half(\rho^{\prime \circ} + \bar \rho^{\prime \circ})\Omega^\circ + \half(\Psi_2^\circ - \bar \Psi_2^\circ)$\\
$\Psi^\circ_2$ & 0 & 0 & 0   \\
\hline
\end{tabular}
\caption{Action of Held's operators on GHP background quantities in type D spacetimes of Kinnersley's Case II, in a tetrad satisfying n2).} \label{tab:1}
\end{table}

Now we consider \eqref{eq:Algspe0}, replacing the
GHP operators in terms of Held's operators \eqref{eq:Hops}, and
expanding all quantities using \eqref{eq:exp}. Each of the resulting
terms is given by a product of integer powers of $\rho$ and $ \bar \rho$
times a quantity annihilated by $\th$.  We then divide by
$\bar \rho^2$ in order to get rid of any positive power of $\bar \rho$,
and for the negative powers of $\bar \rho$ we substitute
\eqref{eq:barrho}. Multiplying by a suitable power $\rho$, the resulting equation is
then of the general form \eqref{eq:poly1}, with coefficients
$a_i^\circ$ built from finite sums and products of
$\Omega^\circ$, $\rho^{\prime \circ}$, $\tau^\circ$, $\Psi_2^\circ$,
$\alpha^\circ_i$, and their Held-derivatives $\Dbar$, $\Dbar'$, and
$\Nbar$. By \eqref{eq:poly1}, we get $a_i^\circ=0$, and this gives a
set of conditions to be satisfied by the 4 coefficient (distributions)
$\alpha^\circ_i, i=0,1,2,3$. 

If we now consider the leading identity $a^\circ_0=0$, or equivalently 
the limit of equation \eqref{eq:Algspe0} at $\sI^+$, then the results of table \ref{tab:1} (and $\rho = -(r-ia\cos \theta)^{-1} \to 0$ 
towards $\sI^+$, with $r \to \infty$ at fixed outgoing Kerr-Newman coordinate $u$, see \eqref{eq:EF}) lead to the equation 
\ben
\label{eq:Algspe1}
0=\{ \Dbar^{\prime 4} \Dbar^4 - 9 (\Psi_2^{\circ})^2_{} \Nbar^2 \} \alpha^\circ_3 
\een
Since $\th = l^a \nabla_a -p\epsilon-q\bar \epsilon$ and $\epsilon=0$ in Kerr in the Kinnersley frame, and since $l^a$ is asymptotically tangent to a congruence of outgoing null geodesics, a GHP quantity annihilated by $\th$ may be viewed geometrically as a function---or rather a section of the appropriate line bundle---on $\sI^+$ 
(or $\sH^-$); see figure~\ref{fig:2}. The outgoing Kerr-Newman coordinates $(u,\theta,\varphi_*)$ as in \eqref{eq:EF} are good coordinates on $\sI^+$ (or $\sH^-$) away from the poles, so as already said, we may think of GHP quantities annihilated by $\th$ in the Kinnersley gauge as 
functions of $(u,\theta,\varphi_*)$. In this frame, the Held operators $\Dbar, \Dbar'$ and $\Nbar$ are 
\ben
\label{eq:Hop}
\begin{split}
\Dbar &= -\frac{1}{\sqrt{2}} \left( \frac{\partial}{\partial \theta} + i \csc \theta \frac{\partial}{\partial \varphi_*} + ia \sin \theta \frac{\partial}{\partial u} 
- \half (p-q) \cot \theta \right), \\
\Dbar' &= -\frac{1}{\sqrt{2}} \left( \frac{\partial}{\partial \theta} - i \csc \theta \frac{\partial}{\partial \varphi_*} - ia \sin \theta \frac{\partial}{\partial u} 
+ \half (p-q) \cot \theta \right), \\
\Nbar &= \frac{\partial}{\partial u} ,
\end{split}
\een
and $\Psi_2^\circ = M$.

We now would like to take the Fourier transform of $ \alpha_3^\circ$ in $u$. In order to be able to do this, we must know the behavior of these functions for $|u| \to \infty$. 
At our disposal, we have our knowledge that the initial data of $h_{ab}$ lie in the space $\sV_0$ (having in particular their support on a Cauchy surface for $r<r_0$), 
as well as \eqref{eq:poly}. Due to the essentially trivial 
dependence upon $r$ of \eqref{eq:poly} which enters only in $\rho=-(r-ia\cos \theta)^{-1}$, we can conclude 
that $\alpha_3^\circ =0$ when $u<u(r_0)$, since $\th^3 (\Psi^{-4/3}_2 \psi_4)=6\alpha_3^\circ$ vanishes for $r>r_0$ on $\Sigma$. 
When $u \to +\infty$, we cannot argue in this manner because $\Psi^{-4/3}_2 \psi_4$ does not necessarily vanish near $r_+$. But we can 
for example use the decay results by \cite{Dafermos:2017yrz} [thm.~10.1, eq.~(233)], see also 
\cite{Ma:2017bxq} [thm.~2],
showing 
that\footnote{Their basic variable is related to ours by $\alpha^{[-2]}=\Psi^{-4/3}_2 \psi_4$.
When considering their energy norms such as $\bar {\mathbb E}_{\tilde \Sigma_\tau}(\alpha^{[-2]})$ appearing in their thm.~10.1, one must take into account that 
$\Delta^{-2}\alpha^{[-2]}$ is smooth at $\sH^+$ in the case considered here, and the relations 
$l^a = (r^2 + a^2)L^a/\Delta , n^a = (r^2+a^2) \underline{L}^a/\Sigma$ between the Kinnersley tetrad and the vector fields $L^a, \underline{L}^a$ employed by these authors.
Furthermore, one should note that on $\sH^+$, $\Delta^{-1}\underline{L}^a$ and $L^a$ have finite limits.
} integrals of the type $I_n(u_2)=\int_{\{r_1<r<r_2, u_1<u<u_2\}} r^n |\Psi^{-4/3}_2 \psi_4|^2 \sin\theta \dd u \dd r \dd \varphi_* \dd \theta$ are uniformly bounded in $u_2$ for fixed $n,u_1$ and some $r_+ < r_1, r_2 < \infty$. It then follows from the special structure of \eqref{eq:poly} (using that $\alpha^\circ_j=\alpha^\circ_j(u,\theta,\varphi_*)$) that if we take a suitable 
linear combination of finitely many $I_n$'s picking out only the contribution $\int_{u_1<u<u_2} |\alpha_3^\circ|^2 \sin\theta \dd u \dd \varphi_* \dd \theta$ after performing the $r$-integrals, that $\alpha^\circ_3$ must be an $L^2$-function in $u$. The mentioned results by \cite{Dafermos:2017yrz,Ma:2017bxq} so far only cover the case that $|a| \ll M$, which is the origin of this restriction in our theorem.

Assuming that this is the case, we can do Fourier-transforms at our hearts content and write
\ben
\label{eq:Fourier}
\alpha_3^\circ(u, \theta, \varphi_*) = \sum_{l,m} \int_{-\infty}^\infty \dd \omega \   {}_{-2} \hat \alpha^\circ_{3, lm}(\omega) \, {}_{-2} S_{\omega lm}(\theta) e^{-i\omega u + im\varphi_*}, 
\een
where ${}_{-2} \hat \alpha^\circ_{3, lm}(\omega)$ is a distribution which is  in $L^2$ with respect to $\omega$, and where ${}_{-2} S_{\omega lm}(\theta)$ are the normalized eigenfunctions 
of the angular spin $s=-2$ Teukolsky master equation \cite{Teukolsky:1973ha, TeukolskyLett1972}. Using either the explicit form \eqref{eq:Hop} and that of the angular Teukolsky equation, 
or going back to the form \eqref{eq:TSI2} and using the identities for the eigenfunctions of the radial Teukolsky equation given e.g., in \cite{Chandrasekhar:1983}, 
one can see that (dropping the subscripts $l,m,\omega$)
\ben
\label{eq:Algspe2}
\{ \Dbar^{\prime 4} \Dbar^4 -9( \Psi_2^{\circ})^2_{} \Nbar^2 \} [S(\theta) e^{-i\omega u + im\varphi_*}] = 
 C^2 \, S(\theta) e^{-i\omega u + im\varphi_*} , 
\een
where the Starobinsky constant $C \equiv {}_{-2} C_{lm}(\omega)$ is given by 
\ben
C^2 = \tfrac{1}{16} \{\lambda^2(\lambda+2)^2 -8\omega^2 \lambda[A^2(5\lambda+6) -12a^2] + 144\omega^2(M^2 + \omega^2 A^4)\},
\een
where $\lambda$ is the eigenvalue of the spheroidal harmonic ${}_{-2} S_{\omega lm}(\theta)$ and where $A=a^2-am/\omega$. Putting together
\eqref{eq:Algspe1} and \eqref{eq:Algspe2}, we therefore learn that 
\ben
{}_{-2} C_{lm}(\omega)^2 \, \hat \alpha^\circ_{3}(\omega) = 0 . 
\een
If\footnote{ In Schwarzschild, the separation constant $\lambda = (l-1)(l+2)$ is independent of $\omega$ and ${}_{-2} C_{lm}(\omega)=0$ manifestly 
has no solutions on the real $\omega$-axis by inspection. The complex solutions correspond to the algebraically special perturbations 
of Schwarzschild found previously by Couch and Newman \cite{Couch:1973zc}.}  
$\omega \mapsto {}_{-2} C_{lm}(\omega)$ does not vanish anywhere on the real axis for any choice of 
$l,m$, then this equation implies $\alpha^\circ_{3}(\omega) = 0$ and then obviously $\alpha_3^\circ=0$ in view of
\eqref{eq:Fourier}. We can then substitute this result into \eqref{eq:Algspe0}, and repeat the same argument, 
leading now to $\alpha_2^\circ=0$, and subsequently to $\alpha_i^\circ=0$ for all $i$.  So, $\psi_4=0$ in view of \eqref{eq:poly}, in addition to $\psi_0=0$. 
Thus, $h_{ab}$ is a perturbation towards another type D metric. These have been analyzed by \cite{Wald:1973jmp} who has shown that $h_{ab}$ is, modulo gauge, either a 
perturbation towards another Kerr metric, or a perturbation towards a rotating C-metric or a NUT metric~\cite{Plebanski:1976gy} (in total 4 real linear independent perturbations). However, 
we are assuming that $h_{ab}$ has vanishing initial data  for $r>r_0$. Thus, such a solution would have vanishing perturbed curvature 
invariants (e.g., $\psi_2$) outside the domain of dependence of the set where the initial data are not zero, which is not the case for the perturbations towards a C- or NUT metric.
So the proof would be complete provided we knew that ${}_{-2} C_{lm}(\omega)$ does not vanish anywhere on the real $\omega$-axis for any choice of $l,m$. 
 
In Kerr, there can be solutions to ${}_{-2} C_{lm}(\omega) = 0$ in the complex $\omega$-plane\footnote{As shown by Chandrasekhar \cite{Chandrasekhar:1983}, such 
frequencies correspond to algebraically special perturbations. The point is that these cannot be in $L^2$.} \cite{Chandrasekhar:1983}, but it is not known to us whether 
there could be any solutions on the real axis.  Whichever way, one can show 
that any solutions to ${}_{-2} C_{lm}(\omega)=0$ on the real axis, or in fact complex plane, must be isolated if they do exist. This follows because 
the separation constant $\lambda$, being obtained from a Sturm-Liouville problem for the angular operator appearing in the separated Teukolsky equation,
with coefficients that are polynomial in $\omega$, must depend analytically on $\omega$.  In other words, 
$\hat \alpha^\circ_3(\omega)$ is a sum of delta-functions and their derivatives peaked at the the countable set $\{\omega_j\} \subset \RR$ where ${}_{-2} C_{lm}(\omega_j)=0$. But this is impossible for an $L^2$ 
function and we conclude again that $\alpha_3^\circ = 0$. Then we proceed in the same manner to successively show that all $\alpha_i^\circ = 0$. The proof is complete. 
\end{proof}

\section{Decomposition of $h_{ab}$ for inhomogeneous equation}
\label{sec:4}

Now we would like to study solutions $h_{ab}$ to the sourced linearized Einstein equation 
\ben
\label{eq:ET}
(\E h)_{ab} = T_{ab} \ , 
\een
where $\E$ is the linearized Einstein operator \eqref{eq:E}. The first part of our analysis will be entirely local and works in any type II
spacetime and with any, possibly distributional, $T_{ab}$. The final part of the analysis involves global features of Kerr, and to 
avoid difficult issues of analytical nature distracting from the main 
points, we will assume that the the source $T_{ab}$ is smooth and of compact support. Of course, for there to be any solutions $h_{ab}$, 
we must have $\nabla^a T_{ab}=0$. 

Ideally, we would like to be able to write the perturbation up to
gauge as $h_{ab} = \Re (\S^\dagger \Phi)_{ab}$ in terms of a Hertz
potential in order to take advantage of the special features of type
II backgrounds. However, it is easy to see that this will not be
possible for a generic $T_{ab}$.  Indeed, such a $h_{ab}$ is
automatically in the ingoing radiation gauge
$h_{nl} = h_{ll} = h_{ml} = h_{m \mb} = 0$, and the $ll$ component of
\eqref{eq:ET} together with \eqref{eqn:Ell_app} for $(\E h)_{ll}$ then
gives $T_{ll}=0$. Thus, if $T_{ll} \neq 0$ somewhere---as happens,
e.g., for the stress tensor of a point particle
\eqref{eq:Tpoint}---then the perturbation can definitely not be
written as $h_{ab} = \Re (\S^\dagger \Phi)_{ab}$ in terms of a Hertz
potential (up to gauge).

The main idea of this paper is that we may nevertheless write $h_{ab} = \Re(\S^\dagger \Phi)_{ab} + x_{ab}$ up to gauge for a relatively simple ``corrector tensor'' $x_{ab}$ determined from $T_{ab}$. Furthermore, the Hertz potential $\Phi$ will be a solution to the adjoint Teukolsky equation with a certain source also determined  from $T_{ab}$. To begin, it is easy to see that without the use of any Einstein equation, a metric perturbation $h_{ab}$ may be put in a gauge where $h_{ab} l^b=0$. Indeed, if $h_{ab}$ is not in this gauge to begin with, we can solve for a gauge vector field $\xi^a$ satisfying 
\ben
(h_{ab} - \Lie_\xi g_{ab}) l^a = 0.
\een
This equation is equivalent to the GHP equations,
\ben
\begin{split}
2 \th \xi_l =& h_{ll},\\
(\th + \bar \rho) \xi_m + (\edth + \tab') \xi_l =& h_{lm},\\
\th \xi_n + \th' \xi_l + (\tau + \tab') \xi_{\mb} + (\tab + \tau') \xi_m =& h_{ln} . 
\end{split}
\label{eq:xidef}
\een
Since $\th = l^a \nabla_a - p\epsilon - q\bar \epsilon$, the first equation is an ordinary differential equation (ODE)
for $\xi_l$ along the orbits of $l^a$. Substituting a solution $\xi_l$ into the second equation, this similarly gives an 
ODE for $\xi_m$ along the orbits of $l^a$. Finally substituting both solutions $\xi_l, \xi_m$ into the third equation 
gives an ODE for $\xi_n$ along the orbits of $l^a$. Having determined the null-tetrad components $(\xi_l, \xi_n, \xi_m, \xi_{\mb})$
of $\xi_a$ we can redefine $h_{ab} \to h_{ab} - {\Lie_\xi g}_{ab}$ in order to obtain a perturbation such that $h_{ab} l^b = 0$. 

The corrector field $x_{ab}$ is now determined in such a way that 
\ben
\label{eq:x}
(T_{ab} - \E x_{ab}) l^b = 0, 
\een
with the idea to eliminate any $l$ component from $T_{ab}$. Our ansatz for $x_{ab}$ is
\ben
\label{eq:xdef}
x_{ab} = 2m_{(a} \mb_{b)} x_{m\mb} -2l_{(a} \mb_{b)} x_{nm} -2l_{(a} m_{b)} x_{n\mb} + l_a l_b x_{nn},
\een
where $x_{n\mb} = \bar x_{nm}$,
so there are 4 real independent components encoded in $x_{m\mb}, x_{mn}, x_{nn}$ by which we attempt to 
satisfy the 4 real independent equations \eqref{eq:x}. We now first transvect \eqref{eq:x} with $l^a$ and use 
the $ll$ component of the Einstein operator $\E$, \eqref{eqn:Ell_app}, to obtain
\ben
\label{eq:xmmb}
\begin{split}
 \{ \th(\th-\rho-\bar \rho) +2\rho\bar\rho\} x_{m\mb} &= T_{ll}
\end{split}
\een
Next, we transvect \eqref{eq:x} with $m^a$ and use 
the $ml$ component of the Einstein operator $\E$, \eqref{eqn:Elm_app}, to obtain:
\ben
\label{eq:xnm}
\begin{split}
& \half\{\th(\th-2\rho) + 2\rhb(\rho-\rhb)\}x_{nm}\\
=&\ T_{lm} - \half\{(\th+\rho-\rhb)(\edth+\tab'-\tau) + 2\tab'(\th-2\rho) -
(\edth-\tau-\tab')\rhb +2\rho\tau\}x_{m\mb}.
\end{split}
\een
Finally, we transvect \eqref{eq:x} with $n^a$ and use 
the $nl$ component of the Einstein operator $\E$, \eqref{eqn:Eln_app}, to obtain:
\ben
\label{eq:xnn}
\begin{split}
 &\half\{\rho(\th-\rho) + \rhb(\th-\rhb)\}x_{nn}\\
 =
 &\ T_{ln}- \half\{(\edthp+\tap-\tab)(\edth-\tau+\tab') +
(\edthp\edth-\tau\tap-\tab\tab'+\tau\tab) - (\Psi_2+{\bar\Psi}_2)\\
&\pheq +(\thp-2\rhp)\rhb + (\th-2\rhb)\rhp +\rho(3\thp-2\rhb')
+\rhb'(3\th-2\rho)\\
&\pheq -2\thp\th + 2\rho\rhb' +2\edthp(\tau)-\tau\tab\}x_{m\mb}
 \\
&\pheq - \half\{(\th-2\rho)(\edthp-\tab) + (\tap+\tab)(\th+\rhb)
-2(\edthp-\tap)\rho-2\tab\th\}x_{nm}\\
&\pheq - \half\{(\th-2\rhb)(\edth-\tau) + (\tab'+\tau)(\th+\rho)
-2(\edth-\tab')\rhb-2\tau\th\}x_{n\mb}.\\
\end{split}
\een
To simplify, we could set $\tau=\tau'=\sigma'=0$ by an appropriate choice of null tetrad; see n1). The system of equations 
\eqref{eq:xmmb}, \eqref{eq:xnm}, \eqref{eq:xnn} is now solved as follows. 
First, since $\th = l^a \nabla_a - p\epsilon - q\bar \epsilon$, the first equation \eqref{eq:xmmb} is an ODE
for $x_{m\mb}$ along the orbits of $l^a$, see figure \ref{fig:2} for the exterior region of Kerr. We take a solution $x_{m\mb}$ and substitute it into the second equation \eqref{eq:xnm}, which thereby becomes an ODE for $x_{nm}$ along the orbits of $l^a$. Finally, we take the solutions $x_{m\mb}, x_{nm}$ and substitute them into the third equation 
\eqref{eq:xnn}, which thereby becomes is an ODE for $x_{nn}$ along the orbits of $l^a$. We can thereby obtain $x_{ab}$ by solving three scalar ODEs, a 
comparatively manageable task in practice.
\usetikzlibrary{decorations.pathmorphing}
\tikzset{zigzag/.style={decorate, decoration=zigzag}}
\begin{figure} 
\begin{center}
\begin{tikzpicture}[scale=0.6, transform shape]
\draw[thick](3,-3)--(0,0)--(3,3);
\draw[blue,directed, ultra thick](1,-1)--(4,2);
\draw[blue, reverse directed, ultra thick](.5,.5)--(3.5,-2.5);
\node[anchor=east] at(2.7,1) {{\Large $l^a$}};
\node[anchor=west] at(2.6,-1) {{\Large $n^a$}};

\draw[double](3,3)--(6,0)--(3,-3);

\node[anchor=west] at(4.3,2.1) {{\Large $\sI^+$}};
\node[anchor=west] at(4.3,-2.1) {{\Large $\sI^-$}};
\node[anchor=west] at(-1.2,0) {{\Large $\sS$}};
\draw (0,0) node[draw,shape=circle,scale=0.4,fill=black]{};
\node[anchor=east] at(1.9,2.1) {{\Large $\sH^+$}};
\node[anchor=east] at(1.9,-2.1) {{\Large $\sH^-$}};

\node[anchor=north] at(3.1,3.8){{\Large $i^+$}};
\node[anchor=west] at(6.2,0) {{\Large $i^0$}};
\node[anchor=south] at(3.1,-3.8){{\Large $i^-$}};

\draw (3,3) node[draw,shape=circle,scale=0.5,fill=white]{};
\draw (6,0) node[draw,shape=circle,scale=0.5,fill=white]{};
\draw (3,-3) node[draw,shape=circle,scale=0.5,fill=white]{};

\end{tikzpicture}
\end{center}
\caption{\label{fig:2}Orbits of the principal null vector field $l^a$ and of $n^a$ in the exterior region of Kerr.}
\end{figure}
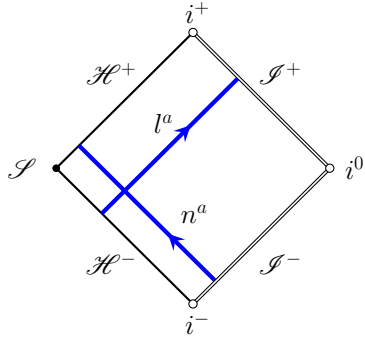

Having determined $x_{ab}$, we now adjust $h_{ab} \to h_{ab} - x_{ab}$. The adjusted $h_{ab}$ still has $h_{ab} l^b = 0$ and additionally satisfies 
 \ben
\label{eq:ES}
(\E h)_{ab} = S_{ab} \equiv T_{ab} - \E x_{ab} \ . 
\een
What $x_{ab}$ has thereby achieved is to remove the $l$ components from the new source $S_{ab}$, i.e., $S_{ab} l^b=0$. At this stage, 
we can apply a result by \cite{Price:2006ke} showing that there exists a gauge vector field $\eta^a$ such that  after a gauge transformation $h_{ab} \to h_{ab} - \Lie_\eta g_{ab}$, the perturbation $h_{ab}$ satisfying \eqref{eq:ES} is in ingoing radiation gauge, $h_{nl} = h_{ll} = h_{ml} = h_{m \mb} = 0$. To summarize, we have obtained from the original perturbation satisfying 
\eqref{eq:ET} by a shift $h_{ab} \to h_{ab} - \Lie_\xi g_{ab} - \Lie_\eta g_{ab} - x_{ab}$ a new perturbation in ingoing radiation gauge satisfying \eqref{eq:ES}. Furthermore, the new source $S_{ab}$ has no $l$ components. Since we have described a constructive procedure to obtain $x_{ab}, S_{ab}$ from the original source $T_{ab}$, we can from now work with the shifted $h_{ab}$ satisfying \eqref{eq:ES}, remembering to add $x_{ab}$ back at the end. 

At this stage, we would like to show that the shifted $h_{ab}$ satisfying \eqref{eq:ES} is of the form $h_{ab} = \Re(\S^\dagger \Phi)_{ab}$, 
for some potential $\Phi \GHPwt (-4,0)$ satisfying the $\O^\dagger$ Teukolsky equation with some source to be determined. From the definition of 
$\S^\dagger$ \eqref{eq:Sdag}, this means that $\Phi$ should simultaneously satisfy the three equations,
\begin{align}
  \label{eq:hmbmb}
  h_{\mb\mb} &= -\half(\th - \rho)(\th + 3\rho) \Phi, \\
  \label{eq:hmbn}
  h_{\mb n} &= -\tfrac{1}{4} \{(\th - \rho + \bar \rho)(\edth + 3\tau) + (\edth-\tau+\bar \tau')(\th +3\rho) \} \Phi, \\
  \label{eq:hnn}
  h_{nn} &= -\half (\edth-\tau)(\edth+3\tau) \Phi + {\rm c.c.}\quad .
\end{align}
It is natural to use the first equation \eqref{eq:hmbmb} in order to define $\Phi$. Since \eqref{eq:hmbmb} can be viewed 
as a second order ODE along the geodesic tangent to $l^a$, integration leaves us with two constants per geodesic. A change of 
those integration constants entails the change
\ben
\label{eq:phich}
\Phi \to \Phi+ A^\circ + \frac{B^\circ}{\rho^3}, 
\een
where $A^\circ \GHPwt (-4,0)$, $B^\circ \GHPwt (-1,3)$, and, as throughout this article, a degree $\circ$ marks a quantity annihilated by $\th$. We shall have to use this freedom momentarily 
when trying to satisfy the second equation, \eqref{eq:hmbn}. In order to understand this relation, we consider the $l\mb$ component of the 
Einstein equation \eqref{eq:ES}, $(\E h)_{l \mb}=0$; see  \eqref{eqn:Elm_app}. Into this equation, we substitute \eqref{eq:hmbmb}, and we use 
the following GHP operator identity:
\ben
\label{eq:GHP1}
\begin{split}
&\half\{\th(\th -2\rhb) +2\rho(\rhb-\rho)\}\{(\th - \rho + \bar \rho)(\edth + 3\tau) + (\edth-\tau+\bar \tau')(\th +3\rho) \}  \\
=& \{(\th-\rho)(\edth-\tab'+\tau) - 2\tau\rho\} (\th - \rho)(\th + 3\rho) ,
\end{split}
\een
which follows from the GHP commutators, Bianchi identity, and background Einstein equation in any type II spacetime\footnote{It is the $l\mb$
component of \eqref{eq:OTdag}.}. Then, if $y$ is defined 
as the left hand side of \eqref{eq:hmbn} minus the right side, we get
\ben
\{\th(\th -2\rhb) +2\rho(\rhb-\rho)\} y = 0.
\een
This is again a second order ODE along the geodesics tangent to $l^a$, which integrates using $\th \rho = \rho^2$ to
\ben
\label{eq:yint}
y=\rhb(\rhb+\rho)a^\circ+\rhb(\rhb-2\rho)\frac{b^\circ}{\rho^3}, 
\een
where $a^\circ \GHPwt (-3,-1),b^\circ \GHPwt (0,2)$ are undetermined GHP scalars annihilated by $\th$. In order to compensate these, we now use \eqref{eq:phich}, under 
which $y$ changes as 
\ben
\label{eq:ych}
y \to y + \tfrac{1}{4} \{(\th - \rho + \bar \rho)(\edth + 3\tau) + (\edth-\tau+\bar \tau')(\th +3\rho) \} \left\{ A^\circ + \frac{B^\circ}{\rho^3} \right\} . 
\een
On the right side, we now substitute for the GHP operators, $\edth, \edth'$ and $\th$ Held's operators $\Dbar, \Dbar'$, and $\Nbar$
\eqref{eq:Hops}, 
and we use the table \ref{tab:2} of operations in type II spacetimes and the following definitions/results \eqref{eq:Helddefn} taken from \cite{Held:1975}, taking advantage also of the special tetrad choice described in n1):
\ben
\label{eq:Helddefn}
\begin{split}
\rho'={}& \rhb \rho^{\prime\circ}-\half \rho(\rho+\rhb)\Psi_2^\circ,\\
\kappa'={}& \kappa^{\prime\circ} - \rho \Psi_3^\circ - \half \rho^2 \Dbar'\Psi_2^\circ - \half \rho^3 \Psi_2^\circ \Dbar' \Omega^\circ,\\
\Psi_3 ={}& \rho^2 \Psi_3^\circ + \rho^3 \Dbar' \Psi_2^\circ + \half 3 \rho^4 \Psi_2^\circ \Dbar' \Omega^\circ\\
\Psi_4={}&\rho \Psi_4^\circ + \rho^2 \Dbar' \Psi_3^\circ + \rho^3(\Psi_3^\circ \Dbar' \Omega^\circ + \half \Dbar' \Dbar' \Psi_2^\circ),\\
&+\rho^4(\half 3 \Dbar' \Psi_2^\circ \Dbar' \Omega^\circ + \half \Psi_2 \Dbar \, \Dbar \, \Omega^\circ) + \half 3 \rho^5 \Psi_2^\circ(\Dbar' \Omega^\circ)^2.
\end{split}
\een

\begin{table}
\begin{tabular}{ | c | c | c | c |}
\hline
& $\Nbar$ & $\Dbar$ & $\Dbar'$ \\ \hline \hline
$\rho$ & 
$\rho^2 \rho^{\prime \circ} - \half \rho(\rho \Psi_2^\circ + \bar \rho \bar \Psi_2^\circ)$ &
$0$ &
$\rho^2 \Dbar' \Omega^\circ$ \\
$\Omega^\circ$ & $\rhb^{\prime \circ} - \rho^{\prime \circ}$ & $\Dbar \, \Omega^\circ$ & $\Dbar' \Omega^\circ$\\ 
$\rho^{\prime \circ}$ & $\Dbar \, \kappa^{\prime\circ}$ & $\Dbar \, \rho^{\prime \circ}$ & $-\Psi_3^\circ-\Omega^\circ \kappa^{\prime\circ}$ \\
$\kappa^{\prime\circ}$ & $\Nbar \kappa^{\prime\circ}$ & $\Dbar \, \kappa^{\prime\circ}$ & $-\Psi_4^\circ$\\ 
$\Psi_4^\circ$ & $\Nbar \Psi_4^\circ$ & $\Nbar \Psi_3^\circ$ & $\Dbar' \Psi_4^\circ$ \\
$\Psi_3^\circ$ & $\Dbar \Psi_4^\circ$ & $\Nbar \Psi_2^\circ$ & $\Dbar' \Psi_3^\circ$ \\
$\Psi_2^\circ$ & $\Dbar \Psi_3^\circ$ & $0$ & $\Dbar' \Psi_2^\circ$ \\
\hline
\end{tabular}
\caption{Action of Held's operators on GHP background quantities in type II spacetimes in a tetrad satisfying n1), taken from \cite{Held:1975}.} \label{tab:2}
\end{table}

Then it is found after a calculation that the change \eqref{eq:ych} will set $y$ as in \eqref{eq:yint} to zero, provided 
$A^\circ, B^\circ$ can be chosen so that 
\ben\label{eq:ab}
\Dbar A^\circ = -2 a^\circ, \quad 
\Dbar B^\circ = -2 b^\circ . 
\een
We shall postpone the discussion of these equations until after theorem \ref{thm:2} and assume that they can be satisfied. Then both \eqref{eq:hmbn}, \eqref{eq:hmbmb} hold, and we can turn our attention to the final equation to be satisfied by $\Phi$, \eqref{eq:hnn}. 
We consider the $ln$ component of the 
Einstein equation \eqref{eq:ES}, $(\E h)_{ln}=0$; see  \eqref{eqn:Eln_app}. Into this equation, we substitute \eqref{eq:hmbmb}, 
\eqref{eq:hmbn}, and we use 
the following GHP operator identity:
\ben
\label{eq:GHP2}
\begin{split}
&\{\rho(\th-\rho) + \rhb(\th-\rhb)\}(\edth-\tau)(\edth+3\tau)\\
=&-\{(\th-2\rhb)(\edth-\tau)+(\tab'+\tau)(\th+\rho) -2(\edth-\tab')\rhb -2\tau \th\} \times\\
&\{ (\th - \rho + \bar \rho)(\edth + 3\tau) + (\edth-\tau+\bar \tau')(\th +3\rho)\}+\\
&\{
(\edth-\tau)(\edth-\tab') - \tab(\tau-\tab')
\}(\th - \rho)(\th + 3\rho),
\end{split}
\een
which follows from the GHP commutators, Bianchi identity, and background Einstein equation in any type II spacetime\footnote{It is the $ln$
component of \eqref{eq:OTdag}.}.
Then, if $z$ is defined 
as the left hand side of \eqref{eq:hnn} minus the right side, we get:
\ben
\{\rho(\th-\rho) + \rhb(\th-\rhb)\} z = 0.
\een
This is a first order ODE along the geodesics tangent to $l^a$, which integrates using $\th \rho = \rho^2$ to:
\ben
\label{eq:zint}
z=(\rhb+\rho)c^\circ, 
\een
where $c^\circ \GHPwt (1,1)$ is a real, undetermined GHP scalar annihilated by $\th$. So the remaining task is to eliminate this $c^\circ$ by whatever
freedom we have left. As noted by \cite{Price:2006ke} there exist gauge vector fields $\zeta^a$ such that  after a gauge transformation $h_{ab} \to h_{ab} - \Lie_\zeta g_{ab}$, the ingoing radiation gauge perturbation $h_{ab}$ satisfying \eqref{eq:ES} is still in ingoing radiation gauge, $h_{nl} = h_{ll} = h_{ml} = h_{m \mb} = 0$. These so called ``residual gauge vector fields'' can be characterized as follows \cite{Price:2006ke}. Denoting 
by $\zeta_l, \zeta_n, \zeta_{m}, \zeta_{\mb}$ the covariant tetrad components of $\zeta^a$ in a tetrad satisfying n1), we have
\ben
\label{eq:resid}
\begin{split}
\zeta_l &= \zeta_l^\circ,\\
\zeta_n &= \zeta_n^\circ + \frac{1}{2}(\Psi_2^\circ\rho + \bar \Psi_2 \rhb) \zeta_l^\circ + \frac{1}{2}( \frac{1}{\rho} + \frac{1}{\rhb}) \Nbar \zeta_l^\circ,\\
\zeta_{\mb} &= \zeta_{\mb}^\circ \frac{1}{\rho} - \Dbar' \zeta_l^\circ, 
\end{split}
\een
where $\zeta_l^\circ, \zeta_n^\circ,\zeta_m^\circ$ are GHP quantities annihilated by $\th$ which must satisfy
\ben
\label{eq:residcond}
\begin{split}
\Nbar \zeta_l^\circ &= -\half(\Dbar' \zeta_m^\circ + \Dbar \, \zeta_{\mb}^\circ),\\
\zeta_n^\circ &= \half(\Dbar' \Dbar + \Dbar \,\Dbar' - \rho^{\prime\circ} -  \rhb^{\prime\circ})\zeta^\circ_l - \half(\Dbar' \zeta_m^\circ - \Dbar \,\zeta_{\mb}^\circ) . 
\end{split}
\een
These conditions leave $\zeta_m^\circ$ undetermined, and we shall now use this freedom to remove $c^\circ$.  First, under a residual 
gauge transformation $h_{ab} \to h_{ab} - \Lie_\zeta g_{ab}$, the Hertz potential $\Phi$ as defined above changes to 
\ben
\Phi \to \Phi - (\Dbar'\Dbar' \zeta_l^\circ + \Dbar' \Omega^\circ \zeta_{\mb}^\circ) \frac{1}{\rho}+\Dbar' \zeta_{\mb}^\circ \frac{1}{\rho^2}, 
\een
whereas $h_{nn}$ changes to
\ben
\label{eq:resid}
\begin{split}
h_{nn} \to h_{nn} &- 2\Nbar \zeta^\circ_n-\Nbar\{(\Psi_2^\circ \rho + \bar \Psi_2^\circ \rhb) \zeta^\circ_l  + (\frac{1}{\rho} + \frac{1}{\rhb}) \Nbar \zeta_l^\circ
 \} \\
& + (\Psi_2^\circ \rho^2 + \bar \Psi_2^\circ \rhb^2)\{
\zeta_n^\circ + \frac{1}{2}(\Psi_2^\circ\rho + \bar \Psi_2 \rhb) \zeta_l^\circ + \frac{1}{2}( \frac{1}{\rho} + \frac{1}{\rhb}) \Nbar \zeta_l^\circ
\}.
\end{split}
\een
Now, $z$ is defined 
as the left hand side of \eqref{eq:hnn} minus the right side, and $z$ is related to $c^\circ$ by \eqref{eq:zint}. 
From this, it follows after a lengthy calculation using formulas from table \ref{tab:2} that, provided that we can chose $\zeta_l^\circ, \zeta_n^\circ,\zeta_m^\circ$ so that 
\ben
\label{eq:c}
\begin{split}
c^\circ=& -\tfrac{1}{4}(\Dbar \, \Dbar \, \Dbar' \Dbar' + \Dbar' \Dbar' \Dbar \, \Dbar \,
-2\Dbar \, \Psi_3^\circ -2 \Dbar' \bar \Psi_3^\circ) \zeta^\circ_l 
+(\Omega^\circ \rho^{\prime \circ} - \Psi_2^\circ) \Dbar' \zeta^\circ_m 
- (\Omega^\circ \rhb^{\prime \circ} + \bar \Psi_2^\circ) \Dbar \, \zeta^\circ_{\mb} \\
&-\half\{\Dbar \Omega^\circ \rhb^{\prime \circ} -\Omega^\circ(\bar \Psi_3^\circ  - \Omega^\circ \bar \kappa^{\prime \circ})
-\half \Dbar \bar \Psi_2^\circ
\} \zeta^\circ_{\mb}
+\half\{\Dbar' \Omega^\circ \rho^{\prime \circ} -\Omega^\circ(\Psi_3^\circ + \Omega^\circ  \kappa^{\prime \circ})
+ \half \Dbar' \Psi_2^\circ
\} \zeta^\circ_{m}\\
&- \tfrac{1}{4}\Dbar' \Omega^\circ \Dbar \, \Dbar \, \zeta^\circ_{\mb}
+ \tfrac{1}{4}\Dbar\, \Omega^\circ \Dbar' \Dbar' \zeta^\circ_{m},
\end{split}
\een
then performing the gauge transformation $h_{ab} \to h_{ab} - \Lie_\zeta g_{ab}$ with $\zeta_l^\circ, \zeta_n^\circ,\zeta_m^\circ$ subject to
\eqref{eq:residcond}, will in effect cancel $c^\circ$. Therefore, \eqref{eq:hnn} will be satisfied. If we combine the gauge vector fields 
defined so far into $X^a = \eta^a + \xi^a + \zeta^a$, then we can summarize our findings to this point as follows. 

\begin{theorem}
\label{thm:2}
Suppose $h_{ab}$ satisfies the inhomogeneous linearized Einstein equation $(\E h)_{ab} = T_{ab}$ on a Petrov type II background, and suppose the symmetric tensor 
field $x_{ab}$ is defined as \eqref{eq:xdef}, with tetrad components obtained by successively solving the three ODEs \eqref{eq:xmmb}, 
\eqref{eq:xnm}, \eqref{eq:xnn}. Then there exists a gauge vector field $X^a$ and a GHP scalar $\Phi\GHPwt (-4,0)$ such that 
\ben
\label{eq:decomp2}
h_{ab} = x_{ab} + (\Lie_X g)_{ab} + \Re(\S^\dagger \Phi)_{ab},
\een
provided that we can solve for given $a^\circ, b^\circ, c^\circ$ the equations \eqref{eq:ab}, \eqref{eq:c} subject to \eqref{eq:residcond}.
\end{theorem}

{\bf Remark:} We note that the transport operators appearing on the left side of equations \eqref{eq:xmmb}, 
\eqref{eq:xnm}, \eqref{eq:xnn} may be rewritten in any type II background as 
\begin{subequations}
\label{eq:partint}
\begin{align}
\half\{\rho(\th-\rho)+\rhb(\th-\rhb)\}=& \half (\rho+\rhb)^2\th \frac{1}{\rho+\rhb},\\
\th(\th-\rho-\rhb) +2\rho\rhb =& \rho^2 \th \frac{\rhb}{\rho^3} \th \frac{\rho}{\rhb},\\
\half\{\th(\th-2\rhb)+2\rho(\rhb-\rho)\} =&  \frac{\rhb}{2(\rho+\rhb)} \th (\rho+\rhb)^2\th \frac{1}{\rhb(\rho+\rhb)}.
\end{align}
\end{subequations}
These identities follow from $\th \rho = \rho^2, \th \rhb = \rhb^2$ and facilitate the integration of the transport equations in 
adapted coordinates, see \eqref{eq:tr1}.

\medskip

Now we investigate whether we can indeed solve for given $a^\circ, b^\circ, c^\circ$ the equations \eqref{eq:ab}, \eqref{eq:c} subject to \eqref{eq:residcond}. This will in general depend on the chosen type II background, and on the nature of the source $T_{ab}$. From now on, we assume that $(\sM,g_{ab})$ is the exterior region of the Kerr black hole with $0\le |a| \le M$. We also assume that $T_{ab}$ has compact support and (for simplicity) that it is smooth. Furthermore, we assume that $h_{ab}$ is the retarded solution. Then the support of $h_{ab}$ satisfies 
${\rm supp}(h_{ab}) \subset J^+({\rm supp}(T_{ab}))$ in some gauge (e.g., the Geroch-Xanthopoulos gauge \cite{Geroch:1978ur}),
where $J^+$ is the causal future of a set, see figure \ref{fig:3}. Furthermore, the linearized ADM/Bondi-mass and angular momentum vanish, $\dot M=\dot J=0$.

\usetikzlibrary{decorations.pathmorphing}
\tikzset{zigzag/.style={decorate, decoration=zigzag}}
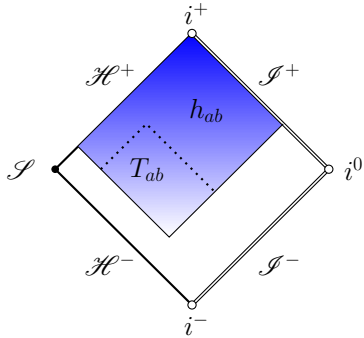
\begin{figure} 
\begin{center}
\begin{tikzpicture}[scale=0.6, transform shape]
\draw[thick](3,-3)--(0,0)--(3,3);

\shade[top color=blue] (.5,.5) -- (3,3) -- (5,1)  -- (2.5,-1.5) -- cycle;
\node[anchor=west] at(1.5,0) {{\Large $T_{ab}$}};
\node[anchor=west] at(2.8,1.3) {{\Large $h_{ab}$}};

\draw[double](3,3)--(6,0)--(3,-3);
\draw(5,1)--(2.5,-1.5)--(.5,.5);
\draw[dotted, thick](1,0)--(2,1)--(3.5,-.5);

\node[anchor=west] at(4.3,2.1) {{\Large $\sI^+$}};
\node[anchor=west] at(4.3,-2.1) {{\Large $\sI^-$}};
\node[anchor=west] at(-1.2,0) {{\Large $\sS$}};
\draw (0,0) node[draw,shape=circle,scale=0.4,fill=black]{};
\node[anchor=east] at(1.9,2.1) {{\Large $\sH^+$}};
\node[anchor=east] at(1.9,-2.1) {{\Large $\sH^-$}};

\node[anchor=north] at(3.1,3.8){{\Large $i^+$}};
\node[anchor=west] at(6.2,0) {{\Large $i^0$}};
\node[anchor=south] at(3.1,-3.8){{\Large $i^-$}};

\draw (3,3) node[draw,shape=circle,scale=0.5,fill=white]{};
\draw (6,0) node[draw,shape=circle,scale=0.5,fill=white]{};
\draw (3,-3) node[draw,shape=circle,scale=0.5,fill=white]{};

\end{tikzpicture}
\end{center}
\caption{
\label{fig:3}
Supports of $T_{ab}$ and $h_{ab}$ in the exterior region of Kerr. 
}
\end{figure}

In Kerr we have the relations $\Nbar \rho^{\prime \circ}=\Nbar \Omega^\circ = \kappa^{\prime \circ} = \Nbar \Psi_2^\circ=
\Dbar \, \Psi_2^\circ=\Psi_3^\circ = 0$ and both $\rho^{\prime \circ}, \Psi_2^\circ$ are real
while $\Omega^\circ$ is by definition imaginary. Furthermore, for any GHP-scalar $x^\circ$ annihilated by $\th$, we have 
$[\Nbar, \Dbar]x^\circ= 0 = [\Nbar, \Dbar']x^\circ$. This leads to significant simplifications in \eqref{eq:c}. Also, 
for any algebraically special spacetime where $\Omega^\circ=0$ such 
as the Robinson-Trautman class or Schwarzschild, \eqref{eq:c} simplifies to
\ben
c^\circ= -\tfrac{1}{4}(\Dbar \, \Dbar \, \Dbar' \Dbar' + \Dbar' \Dbar' \Dbar \, \Dbar \,
-2\Nbar \Psi_2^\circ -2 \Nbar \bar \Psi_2^\circ) \zeta^\circ_l  - (\Psi_2^\circ + \bar \Psi_2^\circ) \Nbar \zeta^\circ_l .
\een
Furthermore, in {\bf Schwarzschild $M>0$}, $\Nbar \Psi_2^\circ = 0, \Psi_2^\circ = M, \Nbar = \partial/\partial u$ in outgoing Kerr-Newman coordinates~\eqref{eq:EF} $(u,r,\theta,\varphi_*)$ and
$c^\circ$ can be viewed as a function on $\sI^+$, i.e., of $(u,\theta,\varphi_*)$, vanishing for $u<u_0$ by construction.
This equation can be solved by noting that it is like a (forward) fourth order heat equation with a source on $\mathbb{R} \times S^2$, with, say, trivial, initial data starting before the source.
The operator replacing the Laplacian in the ordinary heat equation is $\tfrac{1}{4}(\Dbar \, \Dbar \, \Dbar' \Dbar' + \Dbar' \Dbar' \Dbar \, \Dbar \,)$, where in Schwarzschild, $\Dbar, \Dbar'$ are given by \eqref{eq:Hop} with $a=0$.

These operators are equal to the ``spin-weighted'' invariant operators on $S^2$ 
described from a geometric point of view in detail in \cite{eastwood1982edth}. More precisely, a cross section $S^2$ of $\sI^+$ gives rise
to  bundles $\sE^s \to S^2$, where $\sE^s$ is the line bundle isomorphic to an appropriate restriction of $\sL^{p,q}$ to this cross section with ``spin'' $s=\half(p-q)$, and the operators are invariantly defined elliptic operators $\Dbar , \Dbar'$ mapping sections of $\sE^s$ to sections of $\sE^{s+1},\sE^{s-1}$, respectively. From the relation between $\Dbar, \Dbar'$ and the spin-weighted spherical harmonics ${}_{s} Y_{\ell,\mu}$ \cite{Goldberg:1967}, the eigenvalues of this operator are explicitly found to be $\half L(L+1)$ where $L=\ell(\ell+1), \ell=0,1,2,\dots$, and the eigenfunctions are the ${}_{0} Y_{\ell,\mu}$-harmonics. Thus, the ``heat equation'' can be solved by decomposing
into such harmonics. Once we have $\zeta_l^\circ$, the remaining components $\zeta_m^\circ, \zeta_n^\circ$ are obtainable from \eqref{eq:residcond}.

Furthermore, the first equation \eqref{eq:ab} can be solved for $A^\circ$ if and only if $a^\circ \GHPwt (-3,-1)$ is orthogonal to the ${}_{s} Y_{\ell,\mu}$ spin-weighted spherical harmonics with $s=-1,l=1$. Now, the components of the perturbed Bondi angular momentum $\dot J$ of $h_{ab}$ is obtained by taking $r$ times the integral of ${}_{-1} Y_{1,\mu} h_{mn}$ over a sphere $S^2$
at constant $u$ and letting $r \to \infty$, see e.g., \cite{Madler:2016xju}, and we know that $\dot J=0$. From this, it can be seen going through 
the definitions that $a^\circ$ is orthogonal to the  ${}_{-1} Y_{1,\mu},\mu=-1,0,1$
spin-weighted spherical harmonics. On the other hand, one can see that if we choose $\Phi$ as a solution to \eqref{eq:hmbmb} with zero initial 
conditions on $\sH^-$, say, then $b^\circ$ must be zero, because such a term in $y$ as in \eqref{eq:yint} is inconsistent with the behavior \cite{Geroch:1978ur}
of $h_{ab}$ at $\sI^+$ in the Geroch-Xanthopoulos gauge. Thus, it is consistent to take $B^\circ=0$.

\medskip
In {\bf Minkowski, $M=0$} \eqref{eq:c} simplifies to:
\ben
c^\circ =  -\tfrac{1}{4}( \Dbar \, \Dbar \, \Dbar' \Dbar' + \Dbar' \Dbar' \Dbar \, \Dbar) \zeta_l^\circ. 
\een
The operator on the right side is invertible on the subspace orthogonal to the zero mode $\ell=0$. 
The perturbed Bondi-mass of $h_{ab}$ is obtained by taking $r$ times the integral of $h_{nn}$ over a sphere $S^2$
at constant $u$ and letting $r \to \infty$, see e.g., \cite{Madler:2016xju}. Going through the definitions, it follows that if $c^\circ$ is not orthogonal to the constant function, then the perturbation $h_{ab}$ would have a nonzero perturbed ADM (and Bondi-) mass $\dot M$, which cannot be. The equation \eqref{eq:ab} is solved as in Schwarzschild.

In {\bf Kerr,} one can also solve the equations, but the argument is more involved. Thus, we shall now present an alternative argument 
which in effect bypasses equations \eqref{eq:ab} and \eqref{eq:c}. In the argument leading to the preceeding theorem, we go back to the 
definitions of the gauge vector field $\xi^a$ and the tensor $x_{ab}$. We integrate the ODEs required to solve for these quantities, \eqref{eq:xidef}, \eqref{eq:xmmb}, \eqref{eq:xnm}, \eqref{eq:xnn} outwards from $\sH^-$ to $\sI^+$ along the geodesics tangent to $l^a$; see figure \ref{fig:2}. 
Then, due to the support properties of $h_{ab}, T_{ab}$ (see figure \ref{fig:3}) after passing from $h_{ab} \to h_{ab}-(\Lie_\xi g)_{ab}-x_{ab}$, we obtain a new $h_{ab}$ that vanishes in an open neighborhood of $\sH^-$. 
Now, we look at the $ll$ component of the Einstein equation $(\E h)_{ll}=0$ for this new metric. Equation  \eqref{eqn:Ell_app} gives
\ben
0=\{\th(\th-\rho-\rhb)+2\rho\rhb\}h_{m\mb} \equiv (\th-2\rho)(\th+\rho-\rhb)h_{m\mb} , 
\een
which integrates to \cite{Price:2006ke}
\ben
\label{eq:hmmb}
h_{m\mb} = \bar h^\circ \frac{\rho}{\rhb} + h^\circ \frac{\rhb}{\rho} + (k^\circ + \bar k^\circ)(\rho+\rhb),
\een
for undetermined GHP scalars $h^\circ,k^\circ$ annihilated by $\th$. However, $h_{m\mb}=0$ near $\sH^-$. 
By Held's technique we can write \eqref{eq:hmmb} as a polynomial equation in $\rho$ as in \eqref{eq:poly1} with coefficients $a_i^\circ$ annihilated by $\th$. Then these coefficients vanish near $\sH^-$, and therefore all coefficients $a_i^\circ=0$ globally since these quantities are annihilated by $\th$, which is a directional derivative along the orbits of $l^a$ going from $\sH^-$ to $\sI^+$; see figure \ref{fig:2}. One thereby easily finds $h^\circ=k^\circ=0$, i.e., $h_{m\mb}=0$ globally and the metric is automatically in ingoing radiation gauge. The next step in the argument was to integrate the ODE \eqref{eq:hmbmb} for $\Phi$ along the geodesics tangent to $l^a$ going from $\sH^-$ to $\sI^+$. Again, we do this by choosing vanishing initial conditions for 
$\Phi$. Then $y$ as in \eqref{eq:yint} vanishes in a neighborhood of $\sH^-$. By \eqref{eq:barrho}, we can express $y$ as $\rhb^2 \rho^{-3}$ times the polynomial $\Omega^\circ a^\circ \rho^4+2a^\circ \rho^3-\Omega^\circ b^\circ \rho - b^\circ$ in $\rho$ with coefficients annihilated 
by $\th$. Again, by Held's basic lemma \eqref{eq:poly1} $a^\circ, b^\circ$  vanish globally, so $y=0$ globally. Similarly, we show by the same type of argument that the GHP quantity $c^\circ$ appearing \eqref{eq:zint} in fact vanishes automatically, too.

\medskip
 We have shown:

\begin{theorem}
\label{thm:3}
Suppose $h_{ab}$ is a retarded solution to the inhomogeneous linearized Einstein equation $(\E h)_{ab} = T_{ab}$ with smooth compactly 
supported $T_{ab}$ on a Kerr background
having $|a|\le M$, and suppose the symmetric tensor 
field $x_{ab}$ is defined as \eqref{eq:xdef}, with tetrad components obtained by successively solving the three ODEs \eqref{eq:xmmb}, 
\eqref{eq:xnm}, \eqref{eq:xnn}. Then there exists a gauge vector field $X^a$ and a GHP scalar $\Phi\GHPwt (-4,0)$ such that 
\ben
\label{eq:decompi}
h_{ab} = x_{ab} + (\Lie_X g)_{ab} + \Re(\S^\dagger \Phi)_{ab}
\een
\end{theorem}

{\bf Remarks:} 
1) As stated, our result is not symmetric under time-reflection because we have been working with a tetrad satisfying n1) where $l^a$ is a principal null direction while $n^a$ is not (except in Schwarzschild). However, one can derive the same result also for a tetrad satisfying n2) in Kerr, where $l^a$ and $n^a$ are both principal null directions (e.g., the Kinnersley tetrad, see appendix \ref{sec:B}). The derivation proceeds along the same lines, but some formulas, especially those related to the residual gauge vector field $\zeta^a$, given in \cite{Price:2006ke}, are more complicated in this case.

2) Solutions to the inhomogeneous linearized Einstein equation $(\E h)_{ab} = T_{ab}$ with other initial conditions differ from the retarded solution by a solution to the homogeneous equation. We have demonstrated in the previous section that a corresponding decomposition holds to arbitrary precision in a suitable sense, where there can now be an additional piece $\dot g_{ab}$ representing an infinitesimal perturbation towards another Kerr solution.

3) The derivation of our result goes through also for distributional $T_{ab}$ of compact support. 
In that case, $X^a, \Phi$ are distributional as well. 

\medskip

We now turn to the equation satisfied by $\Phi$. To this end, we act with the linearized Einstein operator $\E$ \eqref{eq:E} on \eqref{eq:decompi}, and we use the operator identity \eqref{eq:OTdag}. This results in
\ben \label{eq:86}
\Re(\T^\dagger \O^\dagger \Phi)_{ab} = T_{ab}-(\E x)_{ab} \equiv S_{ab}.
\een
Thus, if $\Phi$ solves the sourced adjoint Teukolsky equation
\ben
\O^\dagger \Phi = \eta
\een
with $\eta$ such that 
\ben
\label{eq:eta}
\Re (\T^\dagger \eta)_{ab} = S_{ab}, 
\een
then $h_{ab}$ as in \eqref{eq:decompi} is a solution to the sourced linearized Einstein equation $(\E h)_{ab} = T_{ab}$, and as discussed, every solution arises in this way. We view \eqref{eq:eta} as the defining equation for $\eta$ and we know that this equation must have solutions. In components, \eqref{eq:eta} is by \eqref{eq:Tdag}
\ben
\label{eq:S1}
S_{\mb \mb} = \tfrac{1}{4} (\th-\rho)(\th-\rho)\eta
\een
\ben
\label{eq:S2}
S_{nn} = \half (\edth -\tau)(\edth-\tau)\eta + {\rm c.c.}
\een
\ben
\label{eq:S3}
S_{\mb n} = \tfrac{1}{4}  \{(\edth + \tab' -\tau)(\th - \rho) + (\th - \rho + \bar \rho)(\edth-\tau) \} \eta.
\een
The left side is in GHP form,
\ben
\label{eq:S1s}
\begin{split}
S_{mm}
=& \pheq T_{mm}-\{(\th-\rhb)(\edth-\tau) - (\edth-\tau-\tab')\rhb -\tau(\th+\rho)
+ \tab'(\th-\rho+\rhb)\}x_{nm} \\
&\pheq -\{(\th-2\rhb)\sib'+ (\tau+\tab')\edth +
(\tau-\tab')^2\}x_{m\mb},
\end{split}
\een
\ben
\begin{split}
S_{nn}
=& \pheq T_{nn}-
\{(\edthp-\tab)(\edth-\tau) +\rhb'(\th-\rho+\rhb) -
(\thp-\rhb')\rhb + {\bar\Psi}_2\}x_{nn}\\
&\pheq - \{-(\thp-3\rhp)(\edthp+\tap-\tab) + \tap\thp - \rhp\edthp -\kap\th\\
&\pheq\pheq +(\th-2\rho+\rhb)\kap + (\edth-3\tau+\tab')\sip +\edth(\sip)
-\Psi_3\}x_{nm}\\
&\pheq - \{-(\thp-3\rhb')(\edth+\tab'-\tau) + \tab'\thp - \rhb'\edth
-\kab'\th\\
&\pheq\pheq +(\th-2\rhb+\rho)\kab' + (\edthp-3\tab+\tap)\sib' +\edthp(\sib') -
{\bar \Psi_3}\}x_{n\mb},
\end{split}
\een

\ben
\begin{split}
S_{n\mb}
=& \pheq T_{n\mb}- \half\{(\th-\rho+\rhb)(\edthp-\tab) - (\edthp-2\tap+\tab)\rho +
\tap(\th-\rhb)\}x_{nn}\\
&\pheq -\half\{-\edthp(\edthp-2\tap)+\sip(\th-2\rho+2\rhb) -
2\tab(\tap-\tab)\}x_{nm}\\
&\pheq -\half\{-(\th+\rhb)(\thp-2\rhb') + \rhp(\th+2\rho-2\rhb) -
4\rho\rhb' + 2\Psi_2\\
&\pheq\pheq + (\edth+\tab')(\edthp-2\tab) - \tap(\edth+\tau-2\tab') -
\tau(\tap-4\tab)\}x_{n\mb}\\
&\pheq -\half\{(\thp+\rhp-\rhb')(\edthp-\tap+\tab) + 2\tab(\thp-2\rhp)
-(\edthp-\tap-\tab)\rhb'+2\rhp\tap \\
&\pheq\pheq+ (\edth-\tau-\tab')\sip + \sip\edth -\kap\th
-\Psi_3\}x_{m\mb},
\end{split}
\een
using the linearized Einstein operator in GHP form given in section \ref{sec:A}.

 $x_{nn},x_{mn},x_{m\mb}$, which are obtained by successively solving in this order the three ODEs \eqref{eq:xmmb}, 
\eqref{eq:xnm}, \eqref{eq:xnn}, can be chosen to vanish near $\sH^-$ if we integrate those equations along the orbits of $l^a$ outwards from $\sH^-$ towards $\sI^+$, 
with vanishing initial conditions, since $T_{ab}$ has compact support. Then also 
$S_{nn},S_{mn}, S_{m\mb}$ vanish near $\sH^-$. \eqref{eq:S1} is an ODE along the geodesics tangent to $l^a$, and we can pick a solution $\eta$ vanishing near $\sH^-$ if we integrate \eqref{eq:S1} outwards from $\sH^-$ towards $\sI^+$ with vanishing initial conditions. Any other solution is of the form $\eta + C^\circ + D^\circ \rho$. We substitute this into the remaining two equations \eqref{eq:S2}, \eqref{eq:S3}, and we apply Held's technique to write the resulting expressions involving $C^\circ, D^\circ$ as a polynomial in $\rho$ as in \eqref{eq:poly1} with coefficients $a_i^\circ$ annihilated by $\th$. Then these polynomials vanish near $\sH^-$, and therefore all coefficients $a_i^\circ=0$ globally since these quantities are annihilated by $\th$, which is a directional derivative along the orbits of $l^a$ going from $\sH^-$ to $\sI^+$. As a consequence, we find writing out the explicit expressions for $a_i^\circ$ in terms of $C^\circ, D^\circ, \Omega^\circ$ and their derivatives by $\Dbar,\Dbar'$ that \eqref{eq:S2}, \eqref{eq:S3} are satisfied if and only if $\Dbar C^\circ = \Dbar D^\circ = 0$, so we may pick $C^\circ=D^\circ =0$. Therefore, the source $\eta$ is given by integrating \eqref{eq:S1} outwards from $\sH^-$ to $\sI^+$ with vanishing initial conditions.

\medskip
 We arrive at the main result of this paper:

\begin{theorem}
\label{thm:4}
Suppose $h_{ab}$ is a retarded solution to the inhomogeneous linearized Einstein equation $(\E h)_{ab} = T_{ab}$ with smooth compactly 
supported $T_{ab}$ on a Kerr background ($ |a| \le M$). Then $h_{ab} = x_{ab} + \Re(\S^\dagger \Phi)_{ab}$ up to gauge, where 
$x_{ab}$ is defined as \eqref{eq:xdef}, with tetrad components obtained by successively solving the three ODEs \eqref{eq:xmmb}, 
\eqref{eq:xnm}, \eqref{eq:xnn} with vanishing initial conditions at $\sH^-$. $\Phi \GHPwt (-4,0)$ is the retarded solution to the adjoint Teukolsky
equation $\O^\dagger \Phi = \eta$. The source $\eta \GHPwt (-4,0)$ is obtained by solving the ODE \eqref{eq:S1} with vanishing initial conditions at $\sH^-$, where $S_{mm}$ is given by \eqref{eq:S1s}.
\end{theorem}

{\bf Remarks:} 
1) As is well known, the adjoint Teukolsky equation can formally be solved by a separation of variables ansatz \cite{TeukolskyLett1972,Teukolsky:1973ha}, i.e., $\Phi \GHPwt (-4,0)$ may be represented in the Kinnersley frame (see appendix \ref{sec:B}) by 
\ben
\Phi(t,r,\theta,\varphi)=\int_{-\infty}^{\infty} {\rm d}\omega \sum_{\ell,m} 
{}_{-2} A_{lm}(\omega) {}_{-2}R_{\omega \ell m}(r)  {}_{-2}S_{\omega \ell m}(\theta)e^{-i\omega t+im\varphi}, 
\een
with radial function ${}_{-2}R_{\omega \ell m}(r)$ determined by the radial spin $-2$ Teukolsky equation \cite{TeukolskyLett1972,Teukolsky:1973ha} with a source ${}_{-2}\eta_{\omega \ell m}(r) $ obtained from $\eta$ by a similar decomposition, and with ${}_{-2}S_{\omega \ell m}(\theta)$ the spin-weighted spheroidal harmonics.
This decomposition can presumably be established rigorously for a retarded solution using recent results by
 \cite{Andersson:2016epf}, who have established that the mode solutions are linearly independent on the real axis. 
 For retarded propagation, as in the statement of the theorem, the radial function satisfies the conditions  
\begin{align}\label{eq:R bcs}
 {}_{-2}R_{\omega \ell m} &\sim r^3 \, e^{i\omega r_*} , \qquad\qquad \  r_*\to+\infty,\nonumber\\
 {}_{-2}R_{\omega \ell m}  &\sim\Delta(r)^2 \, e^{-ikr_*}, \qquad \, r_*\to-\infty,
\end{align}
where $\Delta$ has the conventional meaning for the Kerr metric (see appendix \ref{sec:B}) and
\begin{equation}\label{eq:k def}
k \equiv \omega-m\Omega_H.
\end{equation}
Here $\Omega_H$ is the angular frequency of the outer horizon $\Omega_H=a/(2Mr_+)$. Based on the result by  \cite{Andersson:2016epf}, one should be able to 
prove that standard Laplace-transform expressions based on modes of the retarded propagator for $\O^\dagger$ (as given in, e.g., \cite{Nollert:1999ji}) hold rigorously, thus giving the retarded solution
in mode form.

2) The four ODEs \eqref{eq:xmmb}, 
\eqref{eq:xnm}, \eqref{eq:xnn}, \eqref{eq:S1} defining $x_{ab}$, $\eta$ involve the stress tensor components $T_{ll}$, $T_{nl}$, $T_{ml}$, $T_{mm}$
(in total 6 real components). The remaining 4 real components $T_{nn}, T_{mn}, T_{\mb m}$ enter implicitly through the conservation law 
$\nabla^a T_{ab}=0$ (4 real equations).

3)  Different initial conditions for $h_{ab}$ are encoded in different initial conditions for $\Phi$ plus a perturbation $\dot g_{ab}$ towards another metric in the Kerr family.

4) The derivation of our result goes through also for distributional $T_{ab}$ of compact support inside the chosen exterior region $\sM$ of the Kerr spacetime. In that case, $X^a, \Phi, \eta$ are distributional as well. The compact support assumption about $T_{ab}$ can to some extent be avoided. In fact, our arguments still go through if sufficiently many transverse derivatives of the components $T_{ll}, T_{nl}, T_{ml}, T_{mm}$
on $\sH^-$ vanish and if $T_{ab}$ has sufficiently fast decay towards past infinity for a retarded solution $h_{ab}$ to exist. 

5) By construction, the perturbation $h_{ab} = x_{ab} + \Re(\S^\dagger \Phi)_{ab}$ is in a gauge where it vanishes in an open neighborhood of $\sH^-$. It follows trivially that the perturbed expansion $\dot \vartheta_n$ (i.e., along $n^a$) vanishes on $\sH^-$ and the perturbed expansion 
$\dot \vartheta_l$ vanishes on $\sH^+$ in an open neighborhood of the bifurcation surface, $\sS$ (for $|a| < M$). Then, by the perturbed 
Raychaudhuri equation on $\sH^+$, 
\ben
l^a \nabla_a \dot \vartheta_l = -\vartheta_l \dot \vartheta_l - 2\sigma_l{}^{ab} \dot \sigma_{l \, ab} - T_{ll}, 
\een
since $T_{ll}=0$ on $\sH^+$ and $\vartheta_l=\sigma_{l}^{ab}=0$ on $\sH^+$ in the background, 
$\dot \vartheta_l$ vanishes on the entire future horizon $\sH^+$. As explained in detail in \cite{Hollands:2012sf}, it follows that the perturbation is automatically in a gauge in which the position of $\sH^\pm$ remain in their fixed (background) coordinate location to first order. 

6) All our equations and constructions remain valid if we work with $n^a$, the second repeated principal null direction, instead of $l^a$. The integrations necessary to solve the analogs of the ODEs \eqref{eq:xmmb}, \eqref{eq:xnm}, \eqref{eq:xnn}, \eqref{eq:S1} are now along $n^a$ rather than $l^a$, going from $\sI^-$ to $\sH^+$, rather than from $\sH^-$ to $\sI^+$ (see figure~\ref{fig:2}). Thus, we should prescribe trivial initial conditions at $\sI^-$ for these ODEs rather than at $\sH^-$. The metric obtained in this way will be in the so-called ``outgoing radiation gauge.''

\medskip

Let us finally clarify the relationship between the equation
$\O^\dagger \Phi = \eta$ in theorem~\ref{thm:4} and the inhomogeneous
Teukolsky equation for the perturbed Weyl scalar $\psi_0$ already
given in Teukolsky's original papers~\cite{Teukolsky:1973ha,
  TeukolskyLett1972}. First, we recall how this is derived: applying
the operator relation \eqref{eq:OT} to a symmetric tensor $h_{ab}$
such that $\E h_{ab} = T_{ab}$ gives the relation $\O \psi_0 = T_0$,
where $\psi_0 = \T(h)$ is the perturbed 0-Weyl-scalar
\eqref{eq:Weylcomp} and where $T_0=\S(T)$ [see \eqref{eq:S}] is the source in Teukolsky's
equation for $\psi_0$ \cite{Teukolsky:1973ha, TeukolskyLett1972}.

Now, we substitute for $h_{ab}$ the decomposition
$h_{ab} = x_{ab} + \Re(\S^\dagger \Phi)_{ab} + \Lie_X g_{ab} + \dot g_{ab}$ of
theorem~\ref{thm:4}.\footnote{
Note that in theorem \ref{thm:4}, we are assuming retarded initial conditions. For other initial conditions, 
there would be another piece $\dot g_{ab}$ representing a perturbation towards another Kerr black hole, see remark 2) following thm. \ref{thm:3}, which we can include here without problems.}  Since the perturbed Weyl scalar $\psi_0$ is
gauge invariant on any background in which the corresponding
background scalar $\Psi_0=0$, it follows that $\T(\Lie_X g) =
0$. Furthermore, $\T(x)=0$, since the corrector field $x_{ab}$ has
vanishing $lm,ll,mm$ components [see \eqref{eq:T},\eqref{eq:xdef}]. Finally, 
$\T(\dot g)=0$, since $\Psi_0$ remains zero for perturbations to other Kerr black holes.
Therefore, $\psi_0=\T \Re \S^\dagger \Phi$.  On the other hand, in the
proof of theorem~\ref{thm:4}, we have seen that
$\Re (\T^\dagger \eta)_{ab} = S_{ab} = T_{ab} - (\E x)_{ab}$.
Applying $\S$, as given in \eqref{eq:S}, to this equation and using
$\S\E x = \O \T x = 0$ in view of \eqref{eq:OT}, we find
$T_0=\S \Re \T^\dagger \eta$. The expressions can be further
simplified using \eqref{eq:S}, \eqref{eq:T}, \eqref{eq:Tdag},
\eqref{eq:Sdag} and $\th \rho = \rho^2$, giving
$\S \Re \T^\dagger \eta=-\tfrac{1}{4}
\left(\th^2-4(\rho+\bar\rho)\th+12\rho\bar\rho \right) \th^2 \bar
\eta$ and $\T \Re \S^\dagger \Phi = -\frac{1}{4} \th^4 \bar
\Phi$. Thus, we arrive at the following theorem.

\begin{theorem}
Let $\psi_0$ be the perturbed Weyl scalar associated with a metric perturbation  $h_{ab} = x_{ab} + \Re(\S^\dagger \Phi)_{ab} + \Lie_X g_{ab}$  as in theorem~\ref{thm:4}, with $\Phi$ satisfying $\O^\dagger \Phi = \eta$ with source $\eta$ as described in theorem~\ref{thm:4}. Then 
\ben
\label{eq:Opsi0}
\O \psi_0 = T_0, 
\een
where
\begin{subequations}
\label{eq:etab}
\begin{align}\label{eq:T0 and psi0}
T_0 &= -\tfrac{1}{4} \left(\th^2-4(\rho+\bar\rho)\th+12\rho\bar\rho  \right) \th^2 \bar \eta, \\
\psi_0 &=  -\tfrac{1}{4} \th^4 \bar \Phi.
\end{align}
\end{subequations}
\end{theorem} 

{\bf Remark:} According to this theorem, alternatively, having
obtained $\psi_0$ by solving the Teukolsky equation \eqref{eq:Opsi0}
in terms of $T_0=\S(T)$, one may obtain $\Phi$ by integrating
\eqref{eq:etab} along the orbits of $l^a$. For retarded initial
conditions as in theorem~\ref{thm:4}, we should give zero initial
conditions for this fourth order ODE at $\sH^-$, see
figure~\ref{fig:2}.  Consideration of the Teukolsky equation
$\O \psi_0 = T_0$ alone is nevertheless not sufficient to obtain the
metric perturbation $h_{ab}$: it remains necessary to compute the
corrector tensor $x_{ab}$.

\section{Nonlinear perturbations of Kerr}
\label{sec:5}

So far we have limited the discussion to linearized perturbations. However, even in the absence of matter sources, the linearized metric perturbation  acts as source for the second order perturbation and so on. Merely to count orders---but not implying any specific kind of convergence of the series---let us write 
\begin{equation}
g_{ab} = g_{ab}^{(0)} +  h_{ab}^{(1)}+h_{ab}^{(2)}+ \cdots. 
\end{equation}
$g^{(0)}_{ab}=g^{M,a}_{ab}$ is the background Kerr metric. The equation for $h^{(1)}_{ab}$ is of course just the source free linearized Einstein equation in the absence of matter sources, 
whereas the higher order corrections for $n>1$ satisfy an equation of the form
\ben
\label{eq:En}
(\E h^{(n)})_{ab} = T^{(n)}_{ab}, 
\een
where $\E$ is the linearized Einstein operator in the Kerr-background $g^{M,a}_{ab}$ \eqref{eq:E}, and where the effective stress tensor 
$T^{(n)}_{ab}$ is built out of $h^{(1)}_{ab}, \dots, h^{(n-1)}_{ab}$. For example, the source of  
the equation for $h^{(2)}_{ab}$ is given by minus the nonlinear (quadratic) terms in the second-order  Einstein tensor
\ben
\begin{split}
-T^{(2)}_{cd} = &-\frac12 (\nabla_b h^{(1) ab}-\frac12 g^{ab}\nabla_b h^{(1)})(2\nabla_{(d} h_{c)a)}^{(1)}-\nabla_a h_{ cd}^{(1)}) + 
\frac14 \nabla_c h^{(1) ab} \nabla_d h^{(1)}_{ab} \\
&+ \frac12 \nabla^b h^{(1)a}{}_{c}(\nabla_b h^{(1)}_{ad}-\nabla_a h^{(1)}_{bd}) \\ 
&+ \frac12 h^{(1) ab}(\nabla_c \nabla_d h^{(1)}_{ab}+\nabla_a \nabla_b h^{(1)}_{cd}-2\nabla_{(d} \nabla_{|b|} h^{(1)}_{c)a}).
\end{split}
\een
In the perturbative setting, gauge transformations are formal diffeomorphisms $f={\rm Exp}(\xi^a)$ generated by a formal vector field
\ben
\xi^a = \xi^{(1)a}+ \xi^{(2)a}+\xi^{(3)a}+\cdots
\een
and correspond to $g_{ab} \to f^* g_{ab} = {\rm exp}(\Lie_\xi) g_{ab}$, giving, for instance
\ben
\begin{split}
h^{(1)}_{ab} \to{} & h^{(1)}_{ab} + \Lie_{\xi^{(1)}}g^{(0)}_{ab} \\
h^{(2)}_{ab} \to{} & h^{(2)}_{ab} + \Lie_{\xi^{(2)}}g^{(0)}_{ab} +  \Lie_{\xi^{(1)}} \Lie_{\xi^{(1)}} g^{(0)}_{ab} +\Lie_{\xi^{(1)}}h^{(1)}_{ab} ,
\end{split}
\een
etc. As described in detail in section \ref{sec:3}, at zeroth order, we can apply the decomposition $h^{(1)}_{ab}=g^{(1)}_{ab} + (\Lie_{\xi^{(1)}} g^{(0)})_{ab}+ \Re(\S^\dagger \Phi^{(1)})_{ab}^{}$, where $\Phi^{(1)}$ is a Hertz potential, i.e., a $(-4,0)$ solution to $\O^\dagger \Phi^{(1)}=0$, and where in this section, 
\ben
g^{(n)}_{ab} = \frac{1}{n!} \frac{\dd^n}{\dd s^n} g^{M(s),a(s)}_{ab} 
\een
is an $n$-th order perturbation to another Kerr metric. 

Now, if we could inductively apply at each order $n>1$ the decomposition for $h^{(n)}_{ab}$ described in theorem~\ref{thm:4}  and the following remarks (see section \ref{sec:4}),
with an $n$-th order gauge vector field $\xi^{(n)a}$ and $x_{ab}^{(n)}, \eta^{(n)}$ determined from $T^{(n)}_{ab}$, 
then we could write the metric $g_{ab}$ as the formal series 
\begin{equation}
  \label{eq:decompnonl}
  g_{ab} = 
  g_{ab}^{(0)} + \sum_{n=1}^\infty \left\{g^{(n)}_{ab}  + \Re (\S^\dagger\Phi^{(n)})_{ab} \right\} + \sum_{n=2}^\infty x^{(n)}_{ab}
\end{equation}
with $\O^\dagger \Phi^{(n)} = \eta^{(n)}$, up to a pull back by the
formal diffeomorphism generated by
$\xi^a=\sum_{n=1}^\infty \xi^{(n)a}$.  Since the correctors
$x^{(n)}_{ab}$ and the GHP-sources $\eta^{(n)}$ are comparatively easy
to obtain by solving ordinary transport equations along the orbits of
$l^a$, the analysis of nonlinear perturbations has thereby effectively
been reduced to solving a Teukolsky equation with source at each
order. In view of the powerful methods/results already available in
this setting---such as separation of variables method or the recent
decay results by
\cite{Dafermos:2016uzj,Dafermos:2017yrz,Andersson:2019dwi}---this is
obviously a substantial simplification of matters.

Unfortunately, as stated, theorem~\ref{thm:4} only holds for stress tensors of compact support, while the $n$-th order stress tensors of the gravitational field will clearly not have this property. In order to make our arguments rigorous, we would have to make some approximation of 
$T^{(n)}_{ab}$ by stress tensors of compact support and then take a limit. In order to control the limit, we must understand the asymptotic behavior 
of the perturbations $h^{(1)}_{ab}, \dots, h^{(n-1)}_{ab}$ for large $r$ at finite $u$. These are determined by $\Phi^{(1)},
\dots, \Phi^{(n-1)}$, which are solutions to the Teukolsky equation with a certain source inductively determined, and by $x^{(2)}_{ab}, 
\dots, x^{(n-1)}_{ab}$, obtained by integrating certain transport equations as stated in theorem~\ref{thm:4}. The latter are easy to 
control once we can control in sufficient detail the asymptotic behavior of $\Phi^{(1)},\dots, \Phi^{(n-1)}$. For the Teukolsky equation without source, 
an essentially complete understanding of the asymptotic behavior (in terms of certain energy norms on the initial data) has 
recently been obtained by \cite{Dafermos:2016uzj,Dafermos:2017yrz,Andersson:2019dwi}. We expect that their results, suitably generalized to the Teukolsky equation with source, can be used to show rigorously the decomposition \eqref{eq:decompnonl} to any arbitrary but finite order, thereby also establishing the asymptotic behavior of higher order gravitational perturbations on Kerr (without control over the convergence of the infinite sum \eqref{eq:decompnonl}). However, such an analysis goes beyond the present work and is therefore postponed to another paper.

\section{Summary and outlook}
\label{sec:6}

For the convenience of the reader, we summarize our integration scheme for the sourced linearized Einstein equation $(\E h)_{ab}=T_{ab}$, which consists of the following steps:

{\bf Step 1}: For given $T_{ab}$, we integrate, in this order, the ODEs \eqref{eq:xmmb}, 
\eqref{eq:xnm}, \eqref{eq:xnn} to obtain $x_{m\mb}, x_{nm}, x_{nn}$. 
In the frame \eqref{eq:Kintet up} in outgoing Kerr-Newman coordinates \eqref{eq:EF} $(u,r,\theta, \varphi_*)$, where $\th = \partial/\partial r$, they have the form [using equations~\eqref{eq:partint}]:
\ben
\label{eq:tr1}
\begin{split}
\rho^2 \frac{\partial}{\partial r} \left[ \frac{\rhb}{\rho^3} \frac{\partial}{\partial r} \left(\frac{\rho}{\rhb} x_{m\mb} \right) \right] =& 
{\rm \ r.h.s. \ of \ \eqref{eq:xmmb} } \\
 \frac{\rhb}{2(\rho+\rhb)} \frac{\partial}{\partial r} \left[
 (\rho+\rhb)^2 \frac{\partial}{\partial r} \left( \frac{1}{\rhb(\rho+\rhb)} x_{mn} \right)\right]=&
{\rm \ r.h.s. \ of \ \eqref{eq:xnm} }\\ 
 \frac{1}{2} (\rho+\rhb)^2 \frac{\partial}{\partial r} \left( \frac{1}{\rho+\rhb} x_{nn} \right) =&
 {\rm \ r.h.s. \ of \ \eqref{eq:xnn} }
\end{split}
\een
with right hand sides involving $T_{ll}, T_{ml}, T_{nl}$, respectively and $\rho=-(r-ia\cos\theta)^{-1}$. 
For $x_{m\mb}, x_{nm}, x_{nn}$, we choose 
trivial initial conditions at the (past) event horizon $r=r_+$.

{\bf Step 2}: We integrate the ODE \eqref{eq:S1} in order to obtain $\eta$.
In the frame \eqref{eq:Kintet up} in outgoing Kerr-Newman coordinates \eqref{eq:EF}, this equation has the form:
\ben
\label{eq:tr2}
\begin{split}
\rho \frac{\partial^2}{\partial r^2} \left(\frac{1}{\rho} \eta \right)=
{\rm r.h.s. \ of \ \eqref{eq:S1s} } 
\end{split}
\een
with right hand side involving $T_{mm}, x_{m\mb}, x_{nm}$, respectively. For $\eta$, we choose trivial initial conditions at $r=r_+$.

{\bf Step 3}: Solve the adjoint Teukolsky equation $\O^\dagger \Phi = \eta$ with the desired initial conditions. One may attempt to this end a separation of variables ansatz
\ben
\Phi(u,r,\theta,\varphi_*)=\int_{-\infty}^{\infty} {\rm d}\omega \sum_{\ell,m} {}_{-2} A_{lm}(\omega) {}_{-2}R_{\omega \ell m}(r)  {}_{-2}S_{\omega \ell m}(\theta)e^{-i\omega u+im\varphi_*}, 
\een
with radial function ${}_{-2}R_{\omega \ell m}(r)$ determined by the radial spin $-2$ Teukolsky equation~\cite{TeukolskyLett1972,Teukolsky:1973ha}, and with a source ${}_{-2}\eta_{\omega \ell m}(r) $ obtained from $\eta$ by a similar decomposition. For instance, for a retarded perturbation $h_{ab}$, 
$\Phi$ should be a retarded solution to the adjoint Teukolsky equation, which is equivalent to the boundary conditions \eqref{eq:R bcs} on the modes. In such a case, it might 
be possible to establish the decomposition rigorously if one could justify rigorously the mode decomposition of the retarded propagator based on the standard Laplace-transform method e.g., using recent results by   \cite{Andersson:2016epf}.

{\bf Step 4}: The desired solution of the sourced linearized Einstein equation $(\E h)_{ab}=T_{ab}$ is obtained as 
$h_{ab}=x_{ab}+\Re(\S^\dagger \Phi)_{ab}$, where $x_{ab}$ is constructed from $x_{m\mb}, x_{nm}, x_{nn}$
as in \eqref{eq:xdef}, and where $(\S^\dagger \Phi)_{ab}$ is as in \eqref{eq:Sdag}.

\medskip

By iterating this procedure, then provided theorem 4 can be extended
to noncompact support sources, then one can construct recursively the
$n$-th order gravitational perturbation $h^{(n)}_{ab}$ in terms of an
$n$-th order potential $\Phi^{(n)}$ solving a sourced Teukolsky
equation and an $n$-th order corrector tensor $x^{(n)}_{ab}$.  These
quantities are constructed from the nonlinear terms in the $n$-th
order Einstein tensor, which act as an effective stress tensor
$T^{(n)}_{ab}$ in the $n$-th iteration step.

A different path for constructing higher order (in fact second order) perturbations in a formalism based on the Teukolsky method was proposed some time ago by Campanelli and Lousto \cite{Campanelli:1998jv}. Their starting point is the wave equation for the Weyl tensor, which has the schematic form 
\ben
\nabla^e \nabla_e C_{abcd} = {\rm terms \ quadratic \ in \ } C_{abcd} 
\een
in any Ricci flat ($R_{ab}=0$) spacetime. By transvecting this tensor equation in all possible ways with the legs
$(l^a, n^a, m^a, \mb^a)$ of a Newman-Penrose tetrad, one obtains a system of equations involving the Weyl scalars \eqref{eq:Weylcomp}
and the rotation coefficients (see appendix \ref{sec:B}). By developing the metric $g_{ab}$ around Kerr, one obtains a Teukolsky type equation for the second order perturbed Weyl scalars sourced by terms involving the first order perturbed quantities. To get a closed system of equations, 
one must solve for the first order perturbed rotation coefficients, i.e., effectively the first order perturbed metric, in terms of the first order perturbed (extreme) Weyl scalars. This can be done, e.g., by well-known inversion formulas such as obtained by \cite{Ori:2002uv} at first order. However, the method breaks down as formulated at higher orders, since the inversion formulae no longer apply. It seems that our method, which works directly with the metric perturbation rather than Weyl scalars, is superior in this sense. 

In the future, we would like to use the formalism developed in this paper in combination with the mode projection method outlined in \cite{Green:2019a} to study the interaction of quasinormal modes in the near horizon region of a near-extremal black hole \cite{Green:2019b}. It would also be an interesting project in our view to use our formalism in combination with results obtained by \cite{Dafermos:2016uzj,Dafermos:2017yrz,Andersson:2019dwi} in order to obtain 
decay properties of higher order gravitational perturbations in Kerr. What seems to be required here is primarily a detailed analysis of the decay properties for solutions to the Teukolsky equation with a source (step 3), since the transport equations (steps 1,2) would then be fairly easy to treat. The method developed in this paper can also be applied to obtain the gravitational field of a point particle in the Kerr metric to linear order, and it would be interesting to compare this to the current alternative treatments of this problem. Finally, it might be possible to use our decompositions in order to obtain simplifications for perturbative quantum gravity off a Kerr background.

\medskip
{\bf Acknowledgements:} S.H. is grateful to the Max-Planck Society for supporting the collaboration between MPI-MiS and Leipzig U., grant Proj. Bez. M.FE.A.MATN0003, and to T. Endler for help with a figure. S.R.G. and P.Z. are grateful to Leipzig U. for hospitality during several visits. We would like to thank S. Aksteiner, L. Andersson, S. Gralla, and R. M. Wald for discussions. 

\medskip
{\bf Note:} This is an author-created, un-copyedited version
of an article published in \emph{Classical and Quantum Gravity}. IOP Publishing
Ltd is not responsible for any errors or omissions in this version of
the manuscript or any version derived from it. This article is
published under a \href{https://creativecommons.org/licenses/by/4.0/}{CC BY licence}. The Version of Record is available
online at \url{https://doi.org/10.1088/1361-6382/ab7075}.

\appendix

\section{Linearized Einstein operator in GHP form}
\label{sec:A}

Here we quote from \cite{Price:2006ke} the tetrad components of the linearized Einstein operator $(\E h)_{ab}$ \eqref{eq:E}
in GHP form.\footnote{We thank B. Wardell for pointing out several typos in the expressions in~\cite{Price:2006ke}.} It is assumed that the first null leg $l^a$ of the null tetrad $(l^a, n^a, m^a, \mb^a)$ is 
aligned with a principal null direction of a type II spacetime (so that $\kappa = \sigma = \Psi_0 = \Psi_1 = 0$ in view of the 
Goldberg-Sachs theorem). In a type D spacetime we have additionally the option n2) described in section \ref{sec:2} to set further 
GHP scalars to zero, and in type II we can alternatively make the simplifications described in n1) by an appropriate choice of $n^a$.

\begin{equation}
\begin{split}
(\E h)_{ll} &= \{(\edthp-\tap)(\edth-\tab') + \rho(\thp+\rhp-\rhb') -
(\th-\rho)\rhp + \Psi_2\}h_{ll}\\
&\pheq + \{-(\rho+\rhb)(\th+\rho+\rhb) +4\rho\rhb\}h_{ln}\\
&\pheq + \{-(\th-3\rhb)(\edthp-\tap+\tab) + \tab\th-\rhb\edthp\}h_{lm}\\
&\pheq + \{-(\th-3\rho)(\edth+\tau-\tab') + \tau\th-\rho\edth\}h_{l\mb}\\
&\pheq + \{\th(\th-\rho-\rhb)+2\rho\rhb\}h_{m\mb},
\end{split}
\label{eqn:Ell_app}
\end{equation}

\begin{equation}
\begin{split}
(\E h)_{nn} &= \{2\kap\kab'\}h_{ll}\\
&\pheq + \{(\edthp-\tab)(\edth-\tau) +\rhb'(\th-\rho+\rhb) -
(\thp-\rhb')\rhb + {\bar\Psi}_2\}h_{nn}\\
&\pheq + \{-(\rhp+\rhb')(\thp+\rhp+\rhb') + 4\rhp\rhb'
 - (\edthp-2\tab)\kab' -(\edth-2\tau)\kap\}h_{ln}\\
&\pheq + \{(\thp-\rhb')\kap +\kap(\thp-\rhp-\rhb')
-\kab'\sip\}h_{lm}\\
&\pheq + \{(\thp-\rhp)\kab' +\kab'(\thp-\rhb'-\rhp) -\kap\sib'\}h_{l\mb}\\
&\pheq + \{-(\thp-3\rhp)(\edthp+\tap-\tab) + \tap\thp - \rhp\edthp -\kap\th\\
&\pheq\pheq +(\th-2\rho+\rhb)\kap + (\edth-3\tau+\tab')\sip +\edth(\sip)
-\Psi_3\}h_{nm}\\
&\pheq + \{-(\thp-3\rhb')(\edth+\tab'-\tau) + \tab'\thp - \rhb'\edth
-\kab'\th\\
&\pheq\pheq +(\th-2\rhb+\rho)\kab' + (\edthp-3\tab+\tap)\sib' +\edthp(\sib') -
{\bar \Psi_3}\}h_{n\mb}\\
&\pheq + \{-(\edthp-2\tab)\kap - \sip(\thp-\rhp+\rhb')\}h_{mm} \\
&\pheq + \{-(\edth-2\tau)\kab' - \sib'(\thp-\rhb'+\rhp)\}h_{\mb\mb}\\
&\pheq + \{\thp(\thp-\rhp-\rhb') + \kap(\tau-\tab') +
\kab'(\tab-\tap) +2\sip\sib' + 2\rhp\rhb'\}h_{m\mb},
\end{split}
\end{equation}

\begin{equation}
\begin{split}
(\E h)_{ln} &= \half\{\rhp(\thp-\rhp) + \rhb'(\thp-\rhb') +
(\edth-2\tab')\kap + (\edthp-2\tap)\kab' + 2\sip\sib'\}h_{ll}\\
&\pheq + \half\{\rho(\th-\rho) + \rhb(\th-\rhb)\}h_{nn}\\
&\pheq + \half\{-(\edthp+\tap+\tab)(\edth-\tau-\tab') -
(\edthp\edth+3\tau\tap+3\tab\tab') + 2(\tab+\tap)\edth\\
&\pheq\pheq+ (\th-2\rhb)\rhp +(\thp-2\rhp)\rhb - \rhb'(\th+\rho) -
\rho(\thp+\rhb') -\Psi_2 -{\bar\Psi}_2\}h_{ln}\\
&\pheq + \half\{(\thp-2\rhb')(\edthp-\tap) + \tab(\thp+\rhp+\rhb')
-\tap(\thp-\rhp)\\
&\pheq \pheq-(2\edthp-\tab)\rhb' - (\th-2\rhb)\kap +
\sip(\tau-\tab')\}h_{lm}\\
&\pheq + \half\{(\thp-2\rhp)(\edth-\tab') + \tau(\thp+\rhb'+\rhp)
-\tab'(\thp-\rhb')\\
&\pheq \pheq-(2\edth-\tau)\rhp - (\th-2\rho)\kab' +
\sib'(\tab-\tap)\}h_{l\mb}\\
&\pheq + \half\{(\th-2\rho)(\edthp-\tab) + (\tap+\tab)(\th+\rhb)
-2(\edthp-\tap)\rho-2\tab\th\}h_{nm}\\
&\pheq + \half\{(\th-2\rhb)(\edth-\tau) + (\tab'+\tau)(\th+\rho)
-2(\edth-\tab')\rhb-2\tau\th\}h_{n\mb}\\
&\pheq + \half\{-(\edthp-\tab)(\edthp-\tap) +
\tab(\tab-\tap)-\sip\rho\}h_{mm}\\
&\pheq + \half\{-(\edth-\tau)(\edth-\tab') +
\tau(\tau-\tab')-\sib'\rhb\}h_{\mb\mb}\\
&\pheq + \half\{(\edthp+\tap-\tab)(\edth-\tau+\tab') +
(\edthp\edth-\tau\tap-\tab\tab'+\tau\tab) - (\Psi_2+{\bar\Psi}_2)\\
&\pheq \pheq+(\thp-2\rhp)\rhb + (\th-2\rhb)\rhp +\rho(3\thp-2\rhb')
+\rhb'(3\th-2\rho)\\
&\pheq\pheq -2\thp\th + 2\rho\rhb' +2\edthp(\tau)-\tau\tab\}h_{m\mb},
\end{split}
\label{eqn:Eln_app}
\end{equation}

\begin{equation}
\begin{split}
(\E h)_{lm} &= \half\{(\thp-\rhp)(\edth-\tab')
+(\edth-\tau-2\tab')\rhb' -(\edth-\tau)\rhp +\tau(\thp+\rhp)\\
&\pheq \pheq+\sib'(\edthp-\tap+\tab) + {\bar\Psi}_3+\rhb\kab'\}h_{ll}\\
&\pheq +\half\{-(\th-\rho+\rhb)(\edth+\tau-\tab') -
(\edth-3\tau+\tab')\rhb - 2\rho\tab'\}h_{ln}\\
&\pheq +\half\{-(\thp+\rhb')(\th-2\rhb) + \rho(\thp+2\rhp-2\rhb') -
4\rhp\rhb +2\Psi_2\\
&\pheq\pheq + (\edthp+\tab)(\edth-2\tab') - \tau(\edthp+\tap-2\tab)
-\tap(\tau-4\tab')\}h_{lm}\\
&\pheq + \half\{-\edth(\edth-2\tau) - \sib'(\th+2\rhb-4\rho) -
2\tab'(\tau-\tab')\}h_{l\mb}\\
&\pheq + \half\{\th(\th-2\rho) + 2\rhb(\rho-\rhb)\}h_{nm}\\
&\pheq + \half\{-(\th-\rhb)(\edthp-\tap+\tab) + 2\tab\rhb\}h_{mm}\\
&\pheq + \half\{(\th+\rho-\rhb)(\edth+\tab'-\tau) + 2\tab'(\th-2\rho) -
(\edth-\tau-\tab')\rhb +2\rho\tau\}h_{m\mb},
\end{split}
\label{eqn:Elm_app}
\end{equation}

\begin{equation}
\begin{split}
(\E h)_{n\mb} &= \half\{(\thp-\rhp)\kap + \kap\thp +
\kab'\sip\}h_{ll}\\
&\pheq + \half\{(\th-\rho+\rhb)(\edthp-\tab) - (\edthp-2\tap+\tab)\rho +
\tap(\th-\rhb)\}h_{nn}\\
&\pheq + \half\{(-(\thp-\rhp+\rhb')(\edthp+\tap-\tab) -
(\edthp-3\tap+\tab)\rhb' + (\edth-\tau+\tab')\sip\\
&\pheq \pheq-2\sip\edth - \Psi_3-2\rhp\tab\}h_{ln}\\
&\pheq +\{\sip(\rhp-2\rhb')-\kap(\tap-2\tab)+\half\Psi_4\}h_{lm}\\
&\pheq +\half\{(\thp(\thp-2\rhp) + -\kap(\edth-2\tau+2\tab') +
\kab'(\edthp-4\tap+2\tab)\\
&\pheq \pheq+2\rhb'(\rhp-\rhb')+2\sip\sib'\}h_{l\mb}\\
&\pheq +\half\{-\edthp(\edthp-2\tap)+\sip(\th-2\rho+2\rhb) -
2\tab(\tap-\tab)\}h_{nm}\\
&\pheq +\half\{-(\th+\rhb)(\thp-2\rhb') + \rhp(\th+2\rho-2\rhb) -
4\rho\rhb' + 2\Psi_2\\
&\pheq\pheq + (\edth+\tab')(\edthp-2\tab) - \tap(\edth+\tau-2\tab') -
\tau(\tap-4\tab)\}h_{n\mb}\\
&\pheq +\half\{-(\edthp-\tap)\sip - \sip\edthp\}h_{mm}\\
&\pheq +\half\{-(\thp-\rhb')(\edth-\tau+\tab') +2\tab'\rhb'
-\kab'(\th-2\rho+2\rhb) +\edthp(\sib')-\tab\sib'\}h_{\mb\mb}\\
&\pheq +\half\{(\thp+\rhp-\rhb')(\edthp-\tap+\tab) + 2\tab(\thp-2\rhp)
-(\edthp-\tap-\tab)\rhb'+2\rhp\tap \\
&\pheq\pheq+ (\edth-\tau-\tab')\sip + \sip\edth -\kap\th
-\Psi_3\}h_{m\mb},
\end{split}
\end{equation}

\begin{equation}
\begin{split}
(\E h)_{mm} &= \{(\thp-2\rhp)\sib' +
\kab'(\edth+\tau-\tab')\}h_{ll}\\
&\pheq +\{-\edth(\edth-\tau-\tab') -2\tau\tab' +
\sib'(\rho-\rhb)\}h_{ln}\\
&\pheq +\{(\thp-\rhp)(\edth-\tab') - (\edth-\tau-\tab')\rhp +
\tau(\thp+\rhp-\rhb') -(\th-2\rhb)\kab'\\
&\pheq \pheq-\tab'(\th+\rhb')+\tab\sib' -{\bar\Psi}_3\}h_{lm}\\
&\pheq +\{-(\edth-\tau-\tab')\sib'-\sib'(\edth-\tau)\}h_{l\mb}\\
&\pheq +\{(\th-\rhb)(\edth-\tau) - (\edth-\tau-\tab')\rhb -\tau(\th+\rho)
+ \tab'(\th-\rho+\rhb)\}h_{nm}\\
&\pheq +\{-(\thp-\rhp)(\th-\rhb) + (\edth-\tau)\tap -
\tau(\edthp+\tap-\tab) + \Psi_2\}h_{mm}\\
&\pheq +\{(\th-2\rhb)\sib'+ (\tau+\tab')\edth +
(\tau-\tab')^2\}h_{m\mb},
\end{split}
\end{equation}

\begin{equation}
\begin{split}
(\E h)_{m\mb} &= \half\{\thp(\thp-\rhp-\rhb') + 2\rhp\rhb' +
\kap(\tau-\tab') - \kab'(\tab-\tap) + 2\sip\sib'\}h_{ll}\\
&\pheq +\half\{\th(\th-\rho-\rhb) + 2\rho\rhb\}h_{nn}\\
&\pheq +\half\{-(\thp+\rhp-\rhb')(\th-\rho+\rhb) -\thp(\th+\rho)
+\rho(\thp+\rhp-\rhb') -{\bar\Psi}_2\\
&\pheq\pheq+(\edthp-\tab)(\edth-\tau-\tab') + \edthp\edth -
(\edth-2\tab')\tap -
\tab(2\edth+\tab')\\
&\pheq\pheq-2\tau(\edthp-\tab)+2\tap\tab'+\rhb\rhb'\}h_{ln}\\
&\pheq +\half\{-(\thp-2\rhp)(\edthp-2\tab) + \tab(\thp+2\rhp-2\rhb') +
2(\edth-\tab')\sip - \sip\edth\\
&\pheq \pheq-2\tap\rhb'-2\kap(\rho-\rhb)-\Psi_3\}h_{lm}\\
&\pheq +\half\{-(\thp-2\rhb')(\edth-2\tau) + \tau(\thp+2\rhb'-2\rho') +
2(\edthp-\tau')\sib' - \sib'\edthp\\
&\pheq \pheq-2\tab'\rhp-2\kab'(\rhb-\rho)-{\bar\Psi}_3\}h_{l\mb}\\
&\pheq +\half\{-(\th-2\rhb)(\edthp-2\tap) + \tap(\th-2\rho-2\rhb) -
2\rho\tab+4\tap\rhb\}h_{nm}\\
&\pheq +\half\{-(\th-2\rho)(\edth-2\tab') + \tab'(\th-2\rhb-2\rho) -
2\rhb\tau+4\tab'\rho\}h_{n\mb}\\
&\pheq
+\half\{-\tab(\edthp-\tab)-\tap(\edthp-\tap)-(\th-2\rhb)\sip
\}h_{mm}\\
&\pheq +\half\{-\tau(\edth-\tau)-\tab'(\edth-\tab')-(\th-2\rho)\sib'
\}h_{\mb\mb}\\
&\pheq + \half\{2\thp\th -(\thp-\rhb')\rhb - (\th-\rho)\rhp
-\rho(\thp-\rhp+\rhb') - \rhb'(\th+\rho-\rhb)\\
&\pheq\pheq -(\edthp-2\tap)\tab' + \tau(\edthp+2\tab) -
\tap(\edth-\tab') + \tab(\edth+\tau) -\edthp(\tau)\\
&\pheq\pheq -\Psi_2-{\bar\Psi}_2\}h_{m\mb}.
\end{split}
\end{equation}

\section{GHP and Kerr quantities}
\label{sec:B}

In this work we use a Newman-Penrose  tetrad $(l^a,n^a,m^a,\bar{m}^a)$  wherein
\begin{equation}\label{eq:NP met}
g_{ab} = 2l_{(a}n_{b)}-2m_{(a}\bar m_{b)}.
\end{equation}
We choose the normalization $n_al^a=1$ and $m_a\bar m^a=-1$, corresponding to the ``$-2$'' signature. The Kerr metric may be defined by the Kinnersley tetrad, which is convenient in explicit computations. The covariant Boyer-Lindquist coordinate components in the order $(t,r,\theta,\phi)$ are 
\begin{subequations}
\begin{align}\label{eq:Kintet down}
l_a&=\left(-1,\Sigma/\Delta,0,a\sin^2\theta\right), \\
n_a&=\frac{1}{2}\left(-\Delta/\Sigma,-1,0,a\sin^2\theta \Delta/\Sigma\right),\\
m_a&=\frac{1}{\sqrt{2}(r+ia\cos\theta)}\left( -ia \sin\theta,0,\Sigma, i(r^2+a^2)\sin\theta \right),
\end{align}
\end{subequations}
where 
\begin{equation}\label{eq:BL stuff} 
\Delta=r^2+a^2-2Mr, \qquad \Sigma=r^2+a^2\cos^2\theta.
\end{equation}
The Kinnersley tetrad is regular in the exterior region $\sM$ of Kerr, with the exception of $\sI^-, \sH^+$, and the north and south poles.
Both $l^a$, $n^a$ are repeated principal null directions, so this tetrad satisfies n2). A corresponding tetrad 
satisfying n1) may be defined by the null rotation described in n1). With 
\begin{subequations}
\label{eq:EF}
\begin{align}
\label{eq:rstar}
u=& \ t-r_* \equiv t-r-\frac{r_+^2+a^2}{r_+-r_-}\ln \left(\frac{r-r_+}{r_+}\right) + \frac{r_-^2+a^2}{r_+-r_-}\ln \left(\frac{r-r_-}{r_+}\right),\\
\varphi_*=& \ \varphi-\frac{a}{r_+-r_-}\ln
\frac{r-r_+}{r-r_-}
\end{align}
\end{subequations}
the outgoing Kerr-Newman coordinates, the contravariant components in the 
coordinate system $(u,r,\theta,\varphi_*)$ are
\begin{subequations}
\label{eq:Kintet up}
\begin{align}
l^a&=(0,1,0,0), \\
n^a&=(r^2+a^2,-\Delta/2,0,a)/\Sigma,\\
m^a&=\frac{1}{\sqrt{2}(r+ia\cos\theta)}\left( ia \sin\theta,0,1, i\csc \theta \right).
\end{align}
\end{subequations}
These forms are useful to write out the various transport equations along the orbits of $l^a$ because $\th=\partial/\partial r$ is rather simple, and because the coordinate system $(u,r,\theta,\varphi_*)$ is regular at the past horizon $\sH^-$ (i.e., $r=r_+$). 

The complex spin coefficients with definite GHP weights are given by
\begin{subequations}
\begin{align}
\kappa &=m^a l^b \nabla_b l_a \GHPwt  (3,1) ,\\
\tau &= m^a n^b \nabla_b l_a \GHPwt  (1,-1),\\
\sigma &= m^a m^b \nabla_b l_a \GHPwt  (3,-1), \\
\rho &= m^a  \mb^b \nabla_b l_a  \GHPwt  (1,1),
\end{align}
\end{subequations}
together with their primed counterparts $\kappa',\tau',\sigma',\rho'$ defined by exchanging $l^a \leftrightarrow n^a, m^a \leftrightarrow \mb^a$. 
The remaining 4 complex spin coefficients $\epsilon, \epsilon', \beta, \beta'$ may be read off from \eqref{nabladef}. They do not have 
definite GHP weight (i.e., they are not GHP scalars), but in effect form part 
of the definition of the GHP covariant derivative $\Theta$ \eqref{nabladef}, or equivalently, of $\th,\th',\edth,\edth'$. The explicit values of all spin coefficients in 
the Kinnersley tetrad may be found, e.g., in \cite{Teukolsky:1973ha}.

\bibliography{MyReferences.bib}

\end{document}